# Charge Transport in Semiconducting Carbon Nanotube Networks

Nicolas F. Zorn, Jana Zaumseil

Institute of Physical Chemistry, Universität Heidelberg, D-69120 Heidelberg, Germany.

**Abstract**

Efficient and controlled charge transport in networks of semiconducting single-walled carbon nanotubes is the basis for their application in electronic devices, specifically in field-effect transistors and thermoelectrics. The recent advances in selective growth, purification and sorting of semiconducting and even monochiral carbon nanotubes have enabled field-effect transistors with high carrier mobilities and on/off current ratios that were unthinkable a few years ago. They have also allowed researchers to examine the microscopic interplay of parameters such as nanotube length, density, diameter distribution, carrier density, intentional and unintentional defects, dielectric environment *etc*. and their impact on the macroscopic charge transport properties in a rational and reproducible manner. This review discusses the various models that are considered for charge transport in nanotube networks and the experimental methods to characterize and investigate transport beyond simple conductivity or transistor measurements. Static and dynamic absorption, photoluminescence and electroluminescence spectroscopy as well as scanning probe techniques (*e.g.*, conductive atomic force microscopy, Kelvin probe force microscopy), their unique insights in the distribution of charge carriers in a given nanotube network and the resulting current pathways will be introduced. Finally, recommendations for further optimization of nanotube network devices and a list of remaining challenges are provided.

I. **INTRODUCTION**

With the substantial progress of the purification and sorting of semiconducting single-walled carbon nanotubes (SWNTs)[1-3] their potential for application in electronic and optoelectronic devices has come into sharper focus. Aligned arrays and random networks of purely semiconducting SWNTs are now true competitors to silicon[4,5] and other semiconductors as active layers in high-frequency circuits[5-7] especially on flexible substrates.[8,9] SWNTs exhibit high carrier mobilities, are solution-processable, show radiation hardness[10,11] and have been successfully integrated in complementary circuits for modern microprocessors produced in standard CMOS Fabs.[12,13] However, the continued advancement and optimization of such devices is not only an engineering question. A fundamental understanding of the underlying physics of charge transport in thin films, arrays and networks of nanotubes is required to rationally improve performance at the lowest possible cost. In this review we will give an overview of the current knowledge on the many factors that directly influence both the microscopic and macroscopic charge transport properties of a network of purely semiconducting single-walled carbon nanotubes. Recent experimental data and various theoretical models will be introduced and discussed. Finally, we will give an outlook on what information is still missing and what type of samples or theoretical models are required to fill those gaps. We specifically focus on random networks rather than aligned arrays of SWNTs as they are more easily obtained on a large scale with typical solution-processing methods such as spin-coating or inkjet printing.

SWNTs are hollow cylindrical structures of *sp²*-hybridized carbon that can be described as seamless, rolled-up sheets of graphene. Depending on the roll-up angle and diameter (typically about 0.6 – 2 nm) as determined by the chiral vector ($\vec{C}_h$) and identified by the indices (n,m) (see **Fig. 1a**), single-walled carbon nanotubes can be either metallic or semiconducting. The index pair (n,m) is commonly used to unambiguously name the species or 'chirality' of a nanotube. Due to their large aspect ratio with lengths of hundreds of nm to cm, SWNTs are considered one-dimensional objects. The band structure of SWNTs can be derived from the two-dimensional band structure of graphene by imposing additional periodic boundary conditions given by $\vec{C}_h$ as represented by the cutting lines in the zone-folding approximation (**Fig. 1b**).[14,15] These also lead to the characteristic density-of-states (DOS) of SWNTs with sharp peaks termed van Hove singularities (see **Fig. 1c** for the DOS of two semiconducting (n,m) SWNTs with different diameters). The bandgaps ($E_g$) of semiconducting nanotubes scale inversely with their diameter and range from 1.3 to 0.5 eV in most commercially available raw materials. These bandgaps are similar to or indeed significantly smaller than typical semiconductors such as silicon (bandgap ~1.1 eV), GaAs (~1.4 eV), IGZO (~3.2 eV) as well as many organic semiconductors (0.8 – 2.5 eV). While a single nanotube has a defined electronic structure, there are only few examples of electronic devices based on networks or arrays of only one single type of nanotube (*i.e.*, monochiral), despite the significant advances in nanotube sorting. Usually polydisperse samples consisting of many different semiconducting nanotubes are used that vary in terms of diameter, bandgap, length *etc.* but can also exhibit various degrees of nanotube network density, bundling and defects (for a representative atomic force micrograph of a spin-coated nanotube network see **Fig. 1d**). Controlling and understanding the impact of these different parameters on the charge transport properties of a SWNT network is currently the largest challenge for rational device optimization and electronic applications as we will see in the following sections.

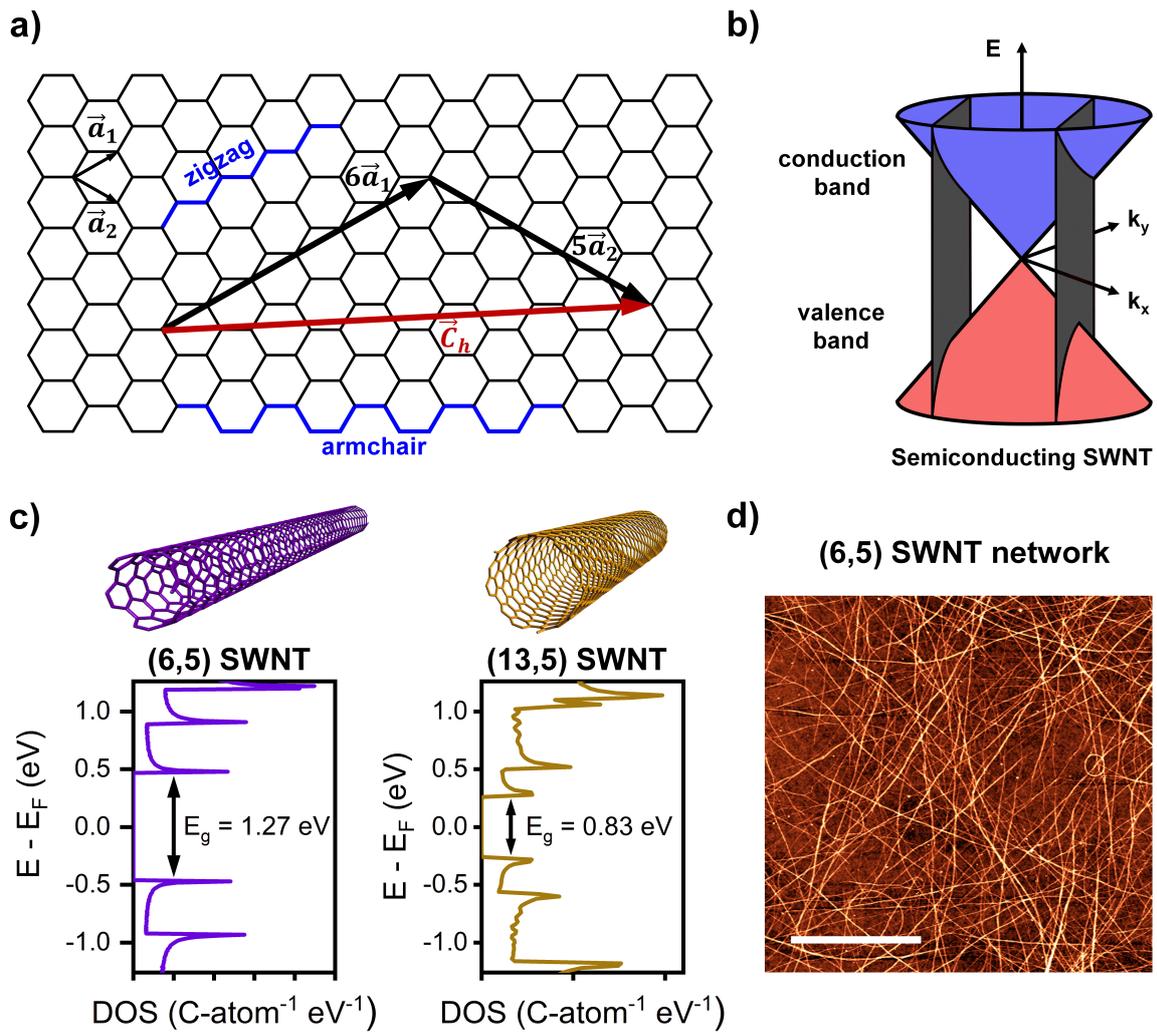

**FIG. 1.** Structural and electronic properties of SWNTs. **(a)** Conceptual construction of a (6,5) SWNT by rolling up a sheet of graphene. The chiral vector $\vec{C}_h$, which forms the circumference of the SWNT, is indicated as a linear combination of $n \cdot \vec{a}_1$ and $m \cdot \vec{a}_2$, with $\vec{a}_1$ and $\vec{a}_2$ being the graphene lattice vectors and (n,m) as the index pair of the nanotube species. **(b)** Zone-folding leads to a quantization of allowed wavevectors and thus one-dimensional carbon nanotube subbands visualized as cuts through the Dirac cone in the band structure of graphene. For semiconducting SWNTs, none of the subbands intersects the K or K' points. **(c)** Illustration of semiconducting (6,5) and (13,5) SWNTs along with their corresponding DOS. (DOS data from Kato, Koretsune, Saito http://www.stat.phys.titech.ac.jp/saito/optCNTs/OptCNT_LDA_DOS_Kato.html). **(d)** Atomic force micrograph of a spin-coated random network of (6,5) SWNTs on SiO$_2$. Scale bar is 1 µm.

## II. DEVICE APPLICATIONS

### A. Field-effect transistors

Field-effect transistors (FETs) are by far the most obvious and successful application of networks of semiconducting SWNTs as only small amounts (about one monolayer) are required to create functional and complex circuits. However, at the same time the requirements for device performance and reproducibility are most stringent and a number of other solution-processable semiconductors with competitive properties already exist (*e.g.*, metal oxides, organic semiconductors).[16,17] Understanding and being able to optimize charge transport in field-effect transistors based on different nanotube networks is crucial for their potential commercialization. Likewise, FETs are very useful tools for the investigation of charge transport at high and controlled carrier densities.

The basic operating principle of a thin film field-effect transistor in accumulation mode - as usually applied for nanotubes - relies on the modulation of the current flow through a semiconducting channel of length $L$ and width $W$, from a source to a drain electrode by a voltage applied to a third electrode - the gate - that is separated from the channel by a thin insulating layer, the gate dielectric. Depending on the areal capacitance ($C_i$ in F/cm$^2$) of the dielectric the applied gate voltage ($V_g$) induces the accumulation of charge carriers at the semiconductor-dielectric interface. For negative gate voltages, holes are injected by the source electrode, whereas for positive gate voltages electrons are injected. As the channel conductance is proportional to the product of the number of charge carriers and their mobility, the increased carrier concentration directly leads to an increased current flow (source-drain current, $I_{ds}$) for a given bias between the drain and the grounded source (source-drain voltage, $V_{ds}$). Consequently, the gate voltage can vary the current flow over many orders of magnitude. Field-effect transistors can be used both as amplifiers and more commonly as electronic switches.

The switching properties of transistors are employed in complex digital circuits,[12,13,18] the simplest being the inverter, but also for turning the pixels in active-matrix displays on and off.[19]

The basic figures of merit for field-effect transistors are the field-effect mobility ($\mu$ in cm$^2$V$^{-1}$s$^{-1}$), the ratio of the drain current in full accumulation (on-state) and depletion (off-state), *i.e.*, the on/off ratio, the subthreshold slope, which indicates how sharply the transistor turns on, and the maximum switching speed, which mainly depends on the carrier mobility, channel length and gate capacitance.[20,21] The specific performance parameters for a given FET can be extracted from current-voltage characteristics such as "transfer curves" ($I_{ds}$ versus $V_g$ for a constant $V_{ds}$) and "output curves" ($I_{ds}$ versus $V_{ds}$ for different constant $V_g$). Specifically, the transconductance $\left(\frac{dI_{ds}}{dV_g}\right)$ of a FET can be determined from transfer curves and used to extract the gate voltage-dependent carrier mobility in the linear (low $V_{ds}$)

$$\mu_{lin} = \frac{dI_{ds}}{dV_g} \cdot \frac{L}{W C_i V_{ds}} \tag{1}$$

or saturation (high $V_{ds}$) regime

$$\mu_{sat} = \frac{dI_{ds,sat}}{dV_g} \cdot \frac{L}{W C_i} \cdot \frac{1}{(V_g - V_{th})} \tag{2}$$

with $V_{th}$ being the threshold voltage for accumulation of mobile charges. Note that these equations are derived based on the gradual channel approximation including a uniform and continuous semiconductor layer. Networks of nanotubes are clearly not completely uniform, in a sparse layer there will be large empty areas. Indeed, the calculated carrier mobility of a SWNT network increases with network density before it saturates.[22,23]

Another unexpected problem for nanotube network transistors is the value for the areal capacitance $C_i$. Typically, this value is calculated from the thickness and dielectric constant of the gate dielectric layer (plate-plate capacitor), however, the one-dimensional nature of nanotubes means that they couple differently to a plate-like gate electrode.[24] Together with the inherent limitation imposed by the quantum capacitance (see below) the actual capacitance of a FET also depends on the network density and is lower than the simple plate-plate capacitance. This may lead to an underestimation of the calculated mobility. The effect can reach two orders of magnitude for thin dielectrics and low network densities.[25] Hence, the capacitance of a nanotube network FET should ideally be measured directly on the device (with an LCR bridge or impedance analyzer)[23] or calculated by taking the network density and dielectric properties into account.[26]

Two typical structures of field-effect transistors, bottom-gate/top-contact and top-gate/bottom-contact, are shown in **Fig. 2a**. The former is often used due to its simplicity (*e.g.* a doped silicon wafer with thermal $SiO_2$ can serve as gate and dielectric, respectively), however, it also suffers from the exposure of the nanotube network to air and the resulting trap states on the polar and hydrophilic $SiO_2$ surface leading to hysteresis and trapping of electrons.[27,28] The latter top-gate structure is more demanding in fabrication but enables self-encapsulation of the active layer and the use of a variety of polymer and oxide dielectrics. The staggered device geometry also helps charge injection of both holes and electrons compared to a coplanar layout as a result of the additional gate field.

Due to the nearly symmetric band structure of carbon nanotubes (at least close to the Fermi level, see **Fig. 1c**), both hole and electron transport are possible with equal effective masses and mobilities, which are indeed observed when electron trap states are removed and no unintentional or intentional doping is applied. Such ambipolar transport is reflected in the transfer characteristics, as shown for a network of purely (6,5) SWNTs in **Fig. 2b**. The electron accumulation (at positive gate voltages) and hole accumulation (at negative gate voltages) branches are nearly symmetrical, and the gap between them is determined by the bandgap of the nanotubes and residual trap states. Due to the absence of metallic nanotubes the off-currents are limited by the gate leakage ($I_g$) and the on/off current ratio reaches $10^6 - 10^7$. The gate voltage dependence of hole and electron mobilities (shown in **Fig. 2c**, extracted by using **Eq. (1)**) with a clear peak is a result of the one-dimensional DOS of SWNTs (see below), but also makes reporting a single mobility value (usually the maximum) more complicated.

The predominant p-type behavior of many SWNT transistors under ambient conditions is the result of the employed high-work function contacts (*e.g.*, gold, palladium or platinum) as well as unintentional p-doping and electron trapping by oxygen and water.[27,28] To obtain n-type transistors, low-work function electron-injecting contacts (*e.g.*, yttrium or scandium)[8,29] are used, while the controlled application of chemical dopants[30] or specific dielectrics[31] are also viable options. An example of a purely n-type nanotube FET after molecular doping, which reduces the work function of the employed gold electrodes, removes electron traps and blocks hole injection[32] is shown in **Fig. 2d**. Both types of unipolar transistors are required for low-power complementary circuits (for the cross-sectional diagram of a SWNT network-based inverter[33] see **Fig. 2e**), which are the basis of modern electronics.

Typical hole and electron field-effect mobilities of random, solution-processed nanotube networks range between 2 and 50 $cm^2V^{-1}s^{-1}$ with some reported values reaching up to 200 $cm^2V^{-1}s^{-1}$.[2,22,34-37] These mobility values are very high for solution-processed semiconductors and sufficient for driving active-matrix organic light-emitting diode displays (AMOLEDs)[19] as well as creating high-frequency circuits as exemplified by GHz-operation of a ring-oscillator based on SWNT network FETs (**Fig. 2f**)[8] or the fabrication of medium-scale nanotube network integrated circuits on a flexible substrate (**Fig. 2g**).[38]

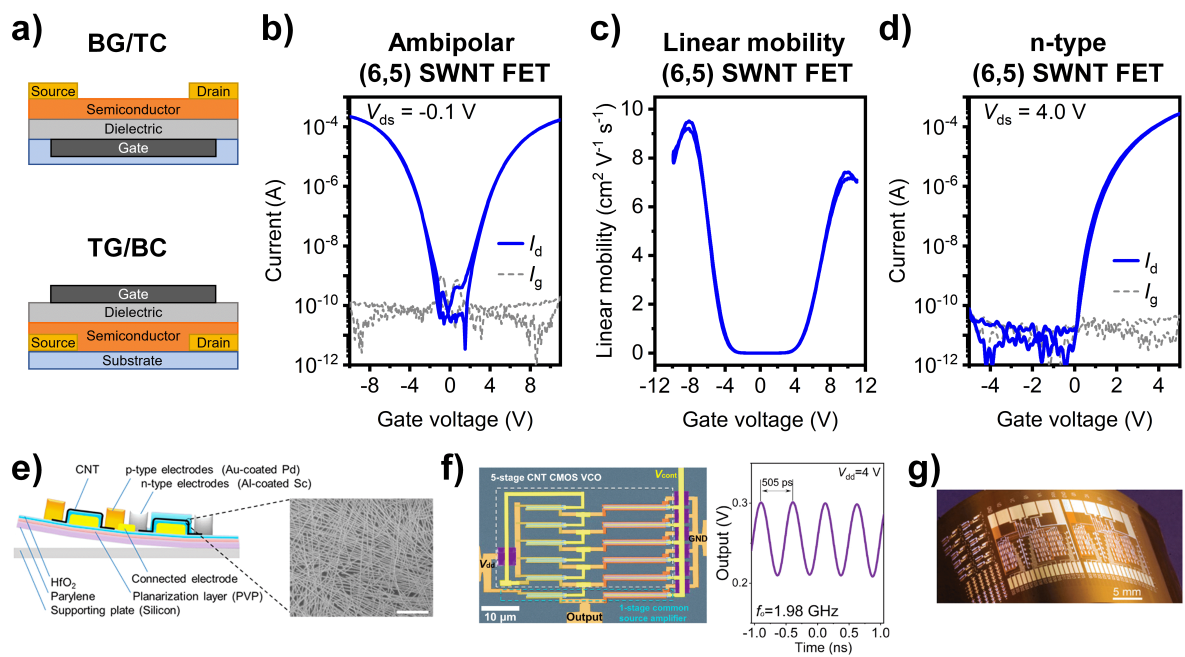

**FIG. 2.** Field-effect transistors and circuits based on SWNTs. **(a)** Schematic illustration of bottom-gate/top-contact (BG/TC) and top-gate/bottom-contact (TG/BC) transistor geometries commonly employed for SWNT network FETs. **(b)** Transfer characteristics of an ambipolar (6,5) SWNT network TG/BC FET and **(c)** corresponding gate voltage-dependent linear mobility. **(d)** Transfer characteristics of a (6,5) SWNT network FET after n-doping with a strong molecular electron donor (ttmgb). Adapted with permission from Schneider *et al.*, *ACS Nano* **2018**, *12*, 5895-5902.[32] Copyright 2018 American Chemical Society. **(e)** Cross-sectional diagram of a SWNT-based inverter structure and scanning electron microscopy (SEM) image of a SWNT film (scale bar, 500 nm). Reproduced with permission from Zhang *et al.*, *ACS Nano* **2018**, *12*, 2773-2779.[33] Copyright 2018 American Chemical Society. **(f)** False-colored SEM image of a five-stage voltage-controlled oscillator (scale bar, 10 μm) based on solution-processed SWNT network FETs and corresponding output voltage signals. Reproduced with permission from Liu *et al.*, *ACS Nano* **2019**, *13*, 2526-2535.[8] Copyright 2019 American Chemical Society. **(g)** Optical image of a SWNT network-based integrated circuit chip on a flexible plastic substrate. Reproduced with permission from Cao *et al.*, *Nature* **2008**, *454*, 495–500.[38] Copyright 2008 Macmillan Publishers Limited.

## B. Other electronic devices

Networks of semiconducting nanotubes have been successfully integrated in a range of optoelectronic devices beyond FETs for circuits. These include light-emitting transistors,[39-42] light-emitting diodes,[43-45] photovoltaic cells,[46-52] photodiodes and -detectors,[53-57] electrochromic cells,[58-62] ratchets,[63] and sensors[64-69] (*e.g.*, for gases, biomolecules, *etc.*). In most of these devices efficient transport of charges is also important but not necessarily the limiting or crucial factor. Also, in some cases the direction of transport is out-of-plane rather than lateral within the network. However, one application, in which the charge transport in a nanotube network also plays an important role is thermoelectrics.

In thermoelectric devices a temperature gradient is turned into a voltage (Seebeck effect) as majority carriers (holes or electrons) move from the hot to the cold side. By connecting p-doped and n-doped semiconducting blocks thermally in parallel and electrically in series an overall current can be source and, *e.g.*, waste heat can be turned into electricity. Numerous reviews are available on the details of novel thermoelectrics, which can be based on a variety of inorganic, organic, composite and nanostructured materials.[70-74] However, with regard to charge transport in carbon nanotube networks it is important to keep in mind that the figure of merit for thermoelectric materials, the *zT* value (unitless), is directly proportional to the electrical conductivity $\sigma$ (S·cm$^{-1}$) and the square of the Seebeck coefficient $\alpha$ (μV·K$^{-1}$) but inversely proportional to the thermal conductivity $\kappa$ (W·m$^{-1}$·K$^{-1}$).

$$zT = \frac{\sigma \alpha^2}{\kappa} T \qquad (3)$$

For maximum power generation from a certain temperature gradient the power factor $\sigma\alpha^2$ has to be maximized. The electrical conductivity obviously depends on the carrier concentration and mobility of the active layer just as in a transistor, although the type and number of mobile carriers is typically determined by chemical doping (p-type or n-type)[75,76] and only in some cases by electrochemical doping (*e.g.*, electrolyte gating).[77,78] The efficiency of the required doping of the nanotube network strongly depends on the combination of nanotube diameter (*i.e.*, bandgap), applied dopants and processing.[76,79,80] The thermal conductivity is governed by lattice phonons and electronic contributions and hence also increases with carrier concentration while the Seebeck coefficient decreases. Furthermore, the thermal conductivity of a single nanotube is exceptionally high with about 3500 W·m$^{-1}$·K$^{-1}$,[81] but the additional phonon scattering by tube–tube junctions and molecular counter-ions of the dopants reduces $\kappa$ in doped networks of SWNTs to values around 1-5 W·m$^{-1}$·K$^{-1}$.[82]

While measuring the precise thermal conductivity of a thin film can be challenging, the Seebeck coefficients depending on hole or electron doping can be determined fairly easily. For that, a linear temperature gradient ($\Delta T$) is created on a substrate through heating at one end and measured with two thermocouples or on-chip resistance thermometers while the thermovoltage ($\Delta V$) between two electrodes is recorded (see **Fig. 3a**) giving the Seebeck coefficient as $\alpha = -\frac{\Delta V}{\Delta T}$. Note that $S_e$ or $S_h$ are also often used as symbols for electron and hole Seebeck coefficients, respectively. Controlled doping of a thin film can be conveniently achieved by electrostatic or electrochemical gating in a transistor structure, which provides the advantage of continuously tuning the carrier density and type (holes or electrons) as shown in **Fig. 3b,c**.[83]

Various theoretical and experimental studies have shown that the Seebeck coefficients of semiconducting nanotubes are higher than those of metallic SWNTs and networks should ideally be composed of purely semiconducting SWNTs.[84,85] Moreover, while the carrier mobility increases with the diameter (see below), the Seebeck coefficient decreases with diameter.[86] Apart from these counteracting fundamental parameters of SWNTs, the morphology of nanotube networks as thermoelectric materials plays an important role. Interestingly, the Seebeck coefficient was found to be the same (isotropic) for parallel and perpendicularly aligned nanotubes, while the carrier mobility and thus conductivity was highly anisotropic (see **Fig. 3b,c**),[83] leading to anisotropic power factors (also for in-plane and out-of-plane of an SWNT film).[75] The complex interplay of these parameters and their dependence on carrier concentration as well as temperature make the development of efficient thermoelectric materials quite challenging. The best $zT$ values of nanotube networks to date are around 0.1, which is still far away from the $zT \approx 1$ of current practical thermoelectric devices.[87] However,

temperature- and carrier density-dependent Seebeck coefficient measurements can be used to gain deeper insights into charge transport in semiconductors and have been applied to nanotube networks as discussed in more detail below.[88]

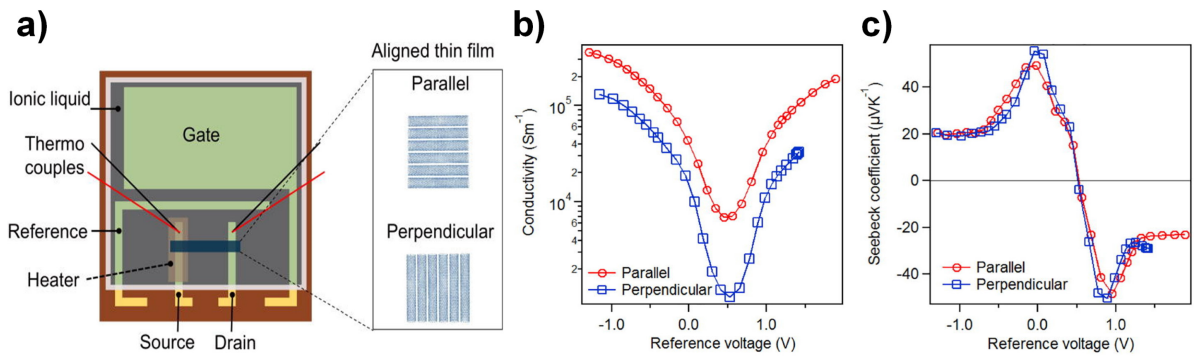

**FIG. 3.** Thermoelectric measurements of nanotubes. (a) Schematic illustration of a setup for measuring the conductivity and Seebeck coefficient of aligned SWNT thin films as a function of applied gate voltage (electron or hole accumulation) in an electrolyte-gated transistor. The temperature gradient is generated by a heater and measured with thermocouples at the source/drain electrodes that also serve to measure thermovoltage and channel conductivity. (b) Conductivities and (c) Seebeck coefficients as a function of gate voltage in the parallel and perpendicular directions of the aligned SWNT thin film. The positive (negative) Seebeck coefficient corresponds to hole (electron) doping. Reprinted with permission from Fukuhara *et al. Appl. Phys. Lett.* **2018**, 113, 243105.[83] Copyright 2018 AIP Publishing.

## III. THEORETICAL MODELS

### A. Percolation in nanotube networks

The overall conductance of any network of conducting wires with a certain length depends on the density and connectivity of these wires as described in general by percolation theory.[89] Below the so-called percolation threshold, the conductance is negligible and around it the conductance increases by orders of magnitude before reaching a nearly constant value. Percolation models (in 3D and 2D) have been especially important for the application of single-walled and multiwalled carbon nanotubes as conductive fillers in insulating polymers[90] but also for transparent conductive electrodes.[91,92] Due to their high aspect ratio the percolation limit for nanotubes in thin films or networks is reached at extremely low concentrations with the precise values depending on the average nanotube length in relation to the electrode distance.[93-95]

With respect to 2D networks of semiconducting nanotubes in field-effect transistors, percolation models were mainly useful for describing the impact of residual metallic nanotubes and the resulting low on/off current ratios.[96-98] Using a simple stick-percolation-based transport model already enabled a better understanding of the influence of nanotube alignment within the channel and the prediction of preferred current pathways.[99-101] For example, Monte Carlo simulations of networks with different levels of nanotube alignment indicated that partially aligned rather than strongly aligned nanotube films should show the lowest resistivity, with the minimum depending on the nanotube length and density as well as channel length.[102] This behaviour was explained with the decreased number of possible junctions and thus conduction pathways for almost perfect alignment (*i.e.*, parallel nanotubes).

Practical nanotube networks are typically far above the percolation threshold, even reaching multilayer thicknesses. Networks with linear densities above 10 SWNTs per micrometer already show no more dependence of the effective carrier mobility on the network density in experiments and thus deviate significantly from the simple 2D percolation model.[23] Nevertheless, more realistic 3D percolation models that could capture the real morphology of a network are still rare[103] and the inclusion of junctions[104] as well as the intrinsic charge carrier density-dependent resistance of different nanotubes poses a significant challenge.

### B. Intra- versus inter-nanotube transport

Assuming the nanotube network is dense enough, such that percolation is not a limiting factor anymore, the question of hole and electron transport in a thin film of only semiconducting carbon nanotubes of certain diameters, lengths *etc.* comes to the fore. **Fig. 4** visualizes the various factors contributing to conductance or resistance in a dense nanotube network: the intrinsic charge transport along a stretch of nanotube, nanotube lengths, nanotube-nanotube junctions (between SWNTs of the same chirality or with different bandgaps), the energetic landscape created by a distribution of different diameters of nanotubes as well as the surrounding dielectric (*e.g.*, trap states or dipolar disorder), scattering due to lattice defects (intentional or unintentional) and field screening in bundles or aggregates. Clearly, the interplay of all of these factors is complex and no analytical model or even numerical simulation exists that can take all of them into account. However, with certain simplifications at least some general trends can be explained.

Let us first consider the intrinsic transport within individual semiconducting carbon nanotubes depending on their specific diameter, band structure and DOS. As shown experimentally and theoretically, individual SWNTs that are longer than the mean free path of electrons (< 1 µm)[15] exhibit diffusive band transport limited by scattering at phonons or possibly defects. This also means that the maximum conductance and charge carrier mobility increase with decreasing temperature due to a lower phonon scattering rate.[105,106] Perebeinos *et al.* found that the impact of electron-phonon scattering on the carrier mobility $\mu$ of an individual (defect-free) nanotube depending on its diameter (*d*) and temperature (*T*) can be described empirically by[105]

$$\mu(T,d) = \mu_0 \frac{300K}{T} \left(\frac{d}{1\,nm}\right)^{\beta} \qquad (4)$$

with $\mu_0$ = 12 000 cm²V⁻¹s⁻¹ and $\beta$ = 2.26. This relation would yield an intrinsic charge carrier mobility of ~6400 cm²V⁻¹s⁻¹ for a single (6,5) SWNT (*d* = 0.757 nm) at room temperature, but a ~3 times higher mobility for a (13,5) SWNT (*d* = 1.278 nm), highlighting the strong impact of the tube diameter on its charge transport properties. Measurements on nanotubes grown by chemical vapour deposition found lower effective mobilities but nearly the same diameter and temperature dependence.[106]

The one-dimensional DOS of nanotubes also results in a limited quantum capacitance[24] and a charge carrier density-dependent mobility with a clear maximum (see **Fig. 2c**). The decrease in carrier mobility after the maximum is due to filling of the first subband, which has been shown theoretically and experimentally for individual nanotubes and nanotube networks.[24,107,108] For even higher carrier concentrations the second subband is filled, as demonstrated for electrolyte-gated small-bandgap nanotubes.[107]

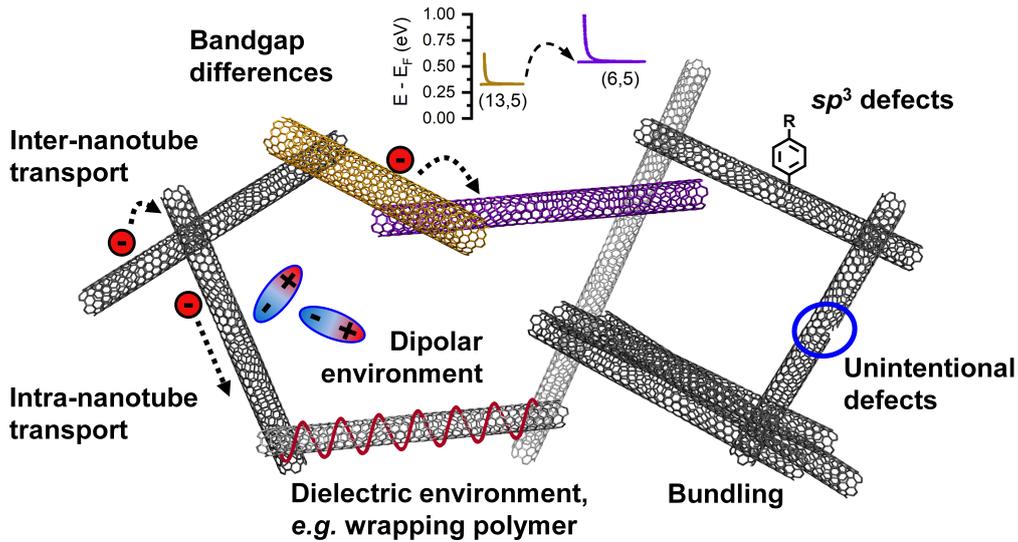

**FIG. 4.** Illustration of different contributions to conductance or resistance in dense networks of purely semiconducting SWNTs: Charge carrier transport across nanotube-nanotube junctions (inter-nanotube) and along nanotube segments (intra-nanotube), bandgap differences, dipolar environment (*e.g.*, at the semiconductor-dielectric interface), dielectric environment (*e.g.*, due to residual wrapping polymer from the nanotube selection process), as well as scattering at unintentional defects (*e.g.*, from growth or processing) and intentional *sp*$^3$ defects.

In SWNT networks, charge carriers do not only move through segments of individual nanotubes but also have to 'hop' across junctions between different nanotubes. These junctions are associated with resistances on the order of $10^2$-$10^5$ k$\Omega$ and have been shown to also depend on the presence of bundles and the electronic type of the crossing nanotubes.[109-111] Tunnelling through a junction is more likely between two nanotubes of large and similar diameter and also depends on the angle between the nanotubes.[112] In a field-effect transistor, realistic junctions are further complicated by the limited gating of a nanotube on top of another or in a bundle due to field screening. A typical assumption is that the junctions are the main bottleneck for transport. A nanotube network can thus be seen as a type of disordered semiconductor or more generally, a disordered material with large conducting regions whose macroscopic conductivity is limited by carrier hopping between the conducting elements. With this simplification one may arrive at an analytical transport model that describes the temperature dependence and ideally also the carrier concentration dependence of the conductivity or apparent carrier mobility. Several models, which originally had been developed to describe charge transport in disordered inorganic or organic semiconductors, have been applied to networks of SWNTs.[113] Most commonly encountered among these models are the variable range hopping (VRH) model[114] and the fluctuation-induced tunnelling (FIT) model.[115]

The VRH model describes the hopping of charges between localized states close to the Fermi level, with the hopping probability depending on the energetic and spatial separation of these states.[20,114] In the framework of the VRH model, the temperature-dependent conductivity $\sigma(T)$ is expressed as

$$\sigma(T) = \sigma_0 \exp\left[-\left(\frac{T_0}{T}\right)^{1/1+d}\right] \quad (5)$$

with the prefactor $\sigma_0$, the dimensionality of the system $d$, and $T_0$ as the hopping parameter. Note that hopping is a thermally activated process and the conductivity and carrier mobility increase with temperature in contrast to the inverse temperature dependence of the intrinsic carrier mobility of SWNTs.

The VRH model can be extended to take into account the carrier density dependence of the observed mobility in monochiral and multichiral carbon nanotube networks. This was demonstrated by Schießl et al. who modelled the SWNT networks as two-dimensional random resistor networks of nanotube-nanotube junctions to emulate their charge transport properties.[108] By including the first subbands of the SWNT density of states, the characteristic carrier density dependence of the simulated network mobilities was in good qualitative agreement with experimental data. Initially, the mobilities increased with carrier density, but decreased again for higher carrier densities.[116] This numerical model could also reproduce the distribution of charges among nanotubes of different diameters in a multichiral semiconducting network.[117]

The FIT model, which was originally developed for conducting polymers, is based on the fluctuation-induced tunnelling of charge carriers through barriers between conductive segments.[115,118] According to the FIT model, the conductivity follows the form described in **Eq. 6**:

$$\sigma(T) = \sigma_0 \exp\left(-\frac{T_1}{T + T_0}\right) \tag{6}$$

where $\sigma_0$ is a temperature-independent prefactor, $T_1$ is the activation energy required to promote a charge carrier over the insulating barrier, and $T_0$ is the temperature above which thermally activated conduction becomes significant. For high temperatures, the FIT model approaches an Arrhenius-type behaviour, at very low temperatures the conductance becomes nearly constant. Again, the conductivity and mobility described in this model increase with temperature. Unfortunately, the FIT model is not easily extended to include the typical DOS of carbon nanotubes.

Modelling entire nanotube networks including intra- and inter-nanotube contributions to charge transport has been a challenging task. Recently, Tripathy *et al.* developed a Monte Carlo model for the field-dependent resistive properties of random SWNT networks (monochiral or multichiral), taking into account the carrier mobilities and resistances of individual nanotubes as well as the inter-nanotube junctions.[104] They found that charge transport in networks is strongly influenced by the density, length and relative orientation of SWNTs, and that phonon scattering is reduced in dense nanotube networks, leading to higher currents.

## IV. EXPERIMENTAL TECHNIQUES

### A. Temperature-dependent transport measurements

The proposed charge transport models for nanotube networks (see above) can be tested against temperature-dependent measurements of charge carrier mobilities and conductivities in field-effect transistors or doped thin films. Measuring the change of carrier mobility with temperature (usually from room temperature to cryogenic temperatures) is indeed a commonly used tool to investigate charge transport in semiconductors and, for example, enables the distinction between band or band-like transport and various types of thermally activated transport in disordered systems. Determining the temperature dependence of charge transport in novel semiconductors such as organic single crystals,[119-122] polymers,[123-126] and inorganic nanomaterials[127-129] has helped to understand their underlying properties and to improve them.

**Thin film conductivity and field-effect transistors**

The temperature and diameter dependence of the transport properties of individual semiconducting SWNTs in back-gated field-effect transistors were studied early on by Zhou *et al.* who found a linear increase in peak mobilities and on-state conductance with $1/T$ in the temperature range between 50 K and 300 K.[106] Furthermore, the maximum mobilities were shown to increase with the square of the nanotube diameter similar to **Eq. 4**. These experimental results supported the theoretical model of diffusive transport limited by scattering of charge carriers with acoustic phonons (see above) and findings for single metallic SWNTs.[105,130]

In SWNT networks, these intrinsic diameter and temperature dependencies should still apply, however, thermally activated charge hopping between different SWNTs may play the more important role. A number of temperature-dependent transport measurements have been conducted on doped and undoped SWNT films,[131-133] as well as unipolar and ambipolar semiconducting SWNT network FETs.[35,118,134-137] Some studies provided evidence that the temperature-dependent resistance of conducting SWNT networks obeyed the VRH model.[131,138,139] However, Yanagi *et al.* demonstrated differences in the conduction behaviour in films of SWNTs depending on the relative metallic/semiconducting nanotube content.[132] In contrast to that, Barnes *et al.* reported that the temperature-dependent conductivity could be fitted with the FIT model even for variable semiconducting-to-metallic SWNT ratios.[133] These early studies reinforced the need for purely semiconducting and well-controlled SWNT networks not only for future device applications, but also for detailed and reliable studies of charge transport.

Itkis *et al.* used temperature-dependent conductance measurements to elucidate charge transport in undoped semiconducting SWNT networks with different thicknesses that were produced by filtration of sorted aqueous nanotube dispersions.[118] They observed that neither of the employed fits (FIT model and variations of the VRH model) could accurately describe the temperature dependence of the conductivity. Pronounced deviations occurred for temperatures above ~70 K (see **Fig. 5a**). Most strikingly, the networks exhibited only a weak modulation of conductivity with temperature compared to conventional semiconductors with a similar bandgap such as germanium. The authors inferred that midgap electronic states created during processing were responsible for the weak temperature dependence of the conductivity.

For a number of different monochiral SWNT films, Gao *et al.* recently observed a pronounced impact of the chirality on the temperature-dependent conductivities.[140] Fitting the data to the VRH model allowed them to determine the carrier localization length, which was found to be significantly higher for metallic (6,6) SWNTs compared to *e.g.* large-bandgap semiconducting (6,5) and (10,3) nanotubes, in agreement with simulated values. However, significant deviations of the experimental data from the VRH model occurred at higher temperatures and especially for metallic armchair (n,n) SWNTs, suggesting a different origin of the temperature dependence for this type of nanotubes.

Shin *et al.* recently performed temperature-dependent electrical measurements on p-type transistors with drop-cast large-diameter nanotube networks.[141] They found a decrease in activation energy for charge transport as determined using the Arrhenius model with decreasing gate voltage. From the transient response of the SWNT FETs and empirically derived rate equations, charge trapping and de-trapping were investigated depending on the temperature, indicating a strong effect on charge transport through the networks.

The temperature-dependent carrier mobilities of bottom-gate FETs based on random networks of polymer-sorted, semiconducting large-diameter SWNTs were first investigated by Gao and Loo.[135] The source-drain currents and corresponding hole mobilities decreased monotonically with decreasing temperature (between 293 K and 78 K), indicating again thermally activated transport. A description with the conventional Arrhenius model proved insufficient, especially for temperatures below 150 K. Instead, the behaviour was best described with the FIT model. In agreement with a study by Kymakis *et al.* on SWNT-polymer composites,[139] it was implied that the wrapping polymer acted as a barrier for inter-tube charge transport in polymer-sorted SWNT networks. The unknown impact of the wrapping polymer on transport, especially when present in excess, is indeed a recurring question and usually considered a problem for comparing and interpreting charge transport measurements in networks based on polymer-sorted SWNTs.[88,134,136,142,143]

Many initial studies on SWNT network field-effect transistors have exclusively employed simple device layouts with two electrodes (source and drain) to extract temperature-dependent mobilities. However, in order to probe the actual carrier mobility, a correction for contact resistance is crucial when the channel resistance is low, as in the case of a high-mobility semiconductor such as a nanotube network. Contact resistance arises at the source and drain electrodes as a result of the formation of a Schottky barrier between metal and nanotube, as shown in various experiments on SWNT FETs with different metallic contacts.[144-146] For commonly employed gold electrodes and large-bandgap (6,5) SWNTs (~ 1.27 eV), the contact resistance-corrected carrier mobilities can be up to 80 % higher compared to the apparent (not corrected) mobilities.[137] The magnitude of this correction, however, depends on the nanotube diameter as the contact resistance is lower for SWNTs with smaller bandgaps. Importantly, the contact resistance itself depends on temperature and thus prevents an unambiguous correlation of temperature-dependent mobility measurements with the intrinsic SWNT network properties.[35] In this context, different device layouts such as gated van der Pauw[147] or gated four-point probe (gFPP) structures[148-150] should be employed. In gFPP devices, the integration of two additional electrodes (voltage probes, usually situated close to the source and drain electrodes) enables the reliable measurement of the actual potential gradient within the channel of an FET when operated in the linear regime. By extrapolation over the entire channel length, the voltage drop at the electrode edges and hence the contact resistance for carrier injection can be determined and the contact-resistance-free mobility can be calculated.[148]

Brohmann *et al.* performed such temperature-dependent gFPP measurements on top-gate FETs with different polymer-sorted SWNT networks.[35] The devices were fabricated from polymer-sorted dispersions of large-diameter semiconducting plasma torch SWNTs (diameter 1.17−1.55 nm), HiPco SWNTs (diameter 0.76−1.31 nm) with a broad bandgap range, and CoMoCAT SWNTs (nearly monochiral (6,5) SWNTs, diameter 0.76 nm) to compare the different network properties and possibly distinguish intra- and inter-nanotube transport. By using a gFPP device layout, the authors were able to extract the contact resistances and the intrinsic field-effect mobilities for temperatures between 100 K and 300 K. A strong dependence of the contact resistance on temperature was found, which underlined the necessity of gFPP method especially for devices with large-bandgap nanotubes.

Decreasing on-currents and on-voltage shifts to higher absolute values for holes and electrons with decreasing temperature were observed for all transistors in this study (see **Fig. 5b** for transfer curves of the monochiral (6,5) SWNTs) with pronounced differences between the different networks. Based on the analysis of the transfer curves, the impact of the mobile carrier concentration on the extracted mobility became clear and thus the problem of comparing mobility values depending on gate voltage. Ideally, either the maximum carrier mobility (if a clear maximum is reached) or the mobility at a certain mobile carrier concentration (constant gate voltage overdrive) should be reported and compared.[35,151]

The temperature dependence of the contact resistance-corrected carrier mobilities for all types of networks in this study was best described by the FIT model (see **Fig. 5c**)[35], although with varying quality and reliability of the fits. The HiPco SWNT network with the largest spread in diameters (bandgaps) exhibited the strongest temperature dependence. However, the relative temperature-dependence of the plasma torch SWNT network consisting of large-diameter nanotubes with a narrow bandgap distribution proved to be consistently lower than for the monochiral (6,5) SWNT network, and the absolute mobilities were higher. While this study focussed on both hole and electron mobilities, the same diameter-dependent trends were later confirmed for doped n-type FETs of the same polymer-sorted nanotube networks.[151] These data indicate that either the junction resistance between nanotubes and its temperature dependence strongly vary with the SWNT diameter or the overall SWNT network mobility is not only governed by the nanotube junctions, but also by the intrinsic temperature ($\mu \sim 1/T$) and diameter dependence ($\mu \sim d^2$) of the intra-nanotube transport.

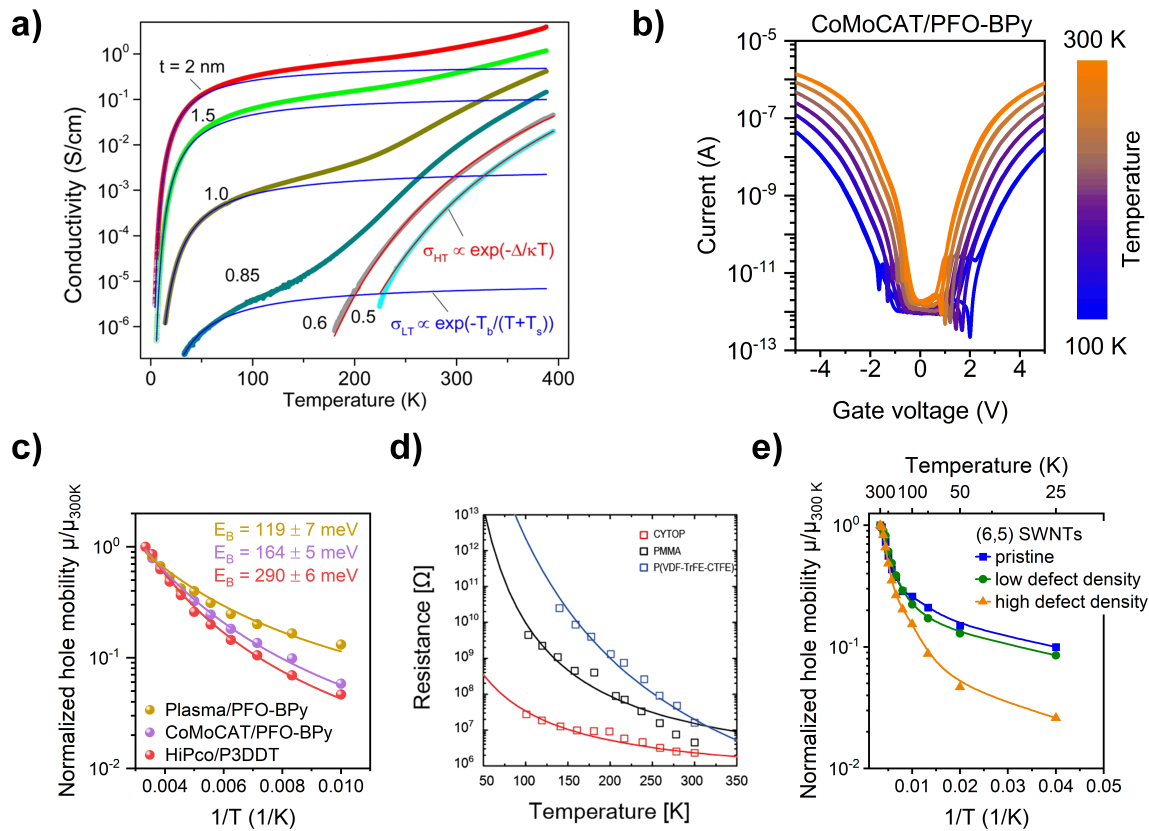

**FIG. 5.** Temperature-dependent transport measurements. **(a)** Temperature-dependent conductivity of semiconducting SWNT networks with different thicknesses (*t*). Reproduced with permission from Itkis *et al*., *Acc. Chem. Res.* **2015**, *48*, 2270–2279.[118] Copyright 2015 American Chemical Society. **(b)** Temperature-dependent ambipolar transfer characteristics (100 K – 300 K) of a FET with a monochiral (6,5) SWNT network. **(c)** Normalized, contact resistance-corrected hole mobilities for FETs based on semiconducting SWNT networks with different diameter and bandgap ranges. Solid lines are fits according to the FIT model. **(b,c)** Adapted with permission from Brohmann *et al*., *J. Phys. Chem. C* **2018**, *122*, 19886–19896.[35] Copyright 2018 American Chemical Society. **(d)** Temperature-dependent resistance of SWNT FETs with different polymeric dielectrics extracted from the p-channel at a fixed carrier concentration. Solid lines represent fits according to the FIT model. Reproduced with permission from Lee *et al*., *ACS Appl. Mater. Interfaces* **2016**, *8*, 32421–32431.[136] Copyright 2016 American Chemical Society. **(e)** Temperature-dependent hole mobilities of pristine and *sp*[3]-functionalized (6,5) SWNT network FETs normalized to the values at 300 K. Mobilities were extracted at a fixed carrier concentration, solid lines are only guides to the eye. Adapted under the terms of the Creative Commons BY-NC-ND 4.0 license from Zorn *et al*., *ACS Nano* **2021**, *15*, 10451-10463.[152] Copyright 2021 American Chemical Society.

This question was further examined in a subsequent study by Brohmann *et al.* where the authors created semiconducting SWNT networks with tailored compositions by mixing large-bandgap, monochiral (6,5) SWNTs and small-bandgap but polydisperse plasma torch SWNTs in different ratios.[137] Temperature-dependent transport measurements revealed the strongest temperature dependence of the contact resistance-corrected carrier mobilities for the (6,5) SWNT network. FETs based on pure plasma torch nanotube networks showed the weakest temperature dependence, but overall behaved quite similar to networks with equal shares of both stock dispersions, which suggested that the latter already contained a substantial number of transport pathways with only large-diameter nanotubes. This systematic difference in temperature-dependence for the different network compositions could again be explained as a superposition of different contributions, namely the thermally activated inter-tube transport across junctions, the dependence of intra-nanotube transport on the inverse temperature, and its dependence on the square of the nanotube diameter.

For many organic semiconductors a strong dependence of the carrier mobility on energetic disorder induced by the randomly oriented dipoles of the gate dielectric reflected in its dielectric constant ($k$) can be found, leading to higher mobilities in FETs with low-$k$ dielectrics such as CYTOP$^{TM}$.[119,153,154] For individual carbon nanotubes the impact of the gate dielectric or polar substrate such as $SiO_2$ has been described theoretically with surface polar phonon scattering, which reduces the intrinsic mobility (limited by acoustic phonon scattering) by an order of magnitude and should indeed be even worse for high-$k$ dielectrics such as $ZrO_2$ or $HfO_2$.[155] Although a large number of mobility values for networks of carbon nanotubes are reported for bottom-gate devices with $SiO_2$ dielectrics, a number of other gate dielectrics ($HfO_2$,[156] $Al_2O_3$,[157] $BaTiO_3$,[158] polymethyl methacrylate (PMMA),[34] styrene–ethylene–butadiene–styrene (SEBS)[159], or even h-BN[160]) and substrates (glass, polyimide (PI), polydimethylsiloxane (PDMS), polyethylene terephthalate (PET) *etc.*)[137,161-163] are nowadays used. Unfortunately, to date there is only one systematic study of the influence of the gate dielectric on charge transport in SWNT FETs.[136] Lee *et al.* reported temperature-dependent measurements (100 K – 300 K) on ambipolar transistors based on polymer-sorted HiPco nanotubes with a very broad diameter distribution (0.8 – 1.2 nm) using PMMA ($k$ = 3.3), CYTOP$^{TM}$ ($k$ = 1.8), and a high-$k$ ferroelectric relaxor P(VDF-TrFE-CTFE) ($k$ = 14.2) as gate dielectrics. They confirmed thermally activated transport in all cases and from fits according to the FIT model, a higher activation energy for transport was obtained for the high-permittivity ferroelectric polymer (see **Fig. 5d**). Correspondingly, PMMA and CYTOP$^{TM}$ exhibited lower activation energies. Similar trends were obtained with Arrhenius fits. However, further systematic studies, especially on monochiral SWNT networks and networks with larger diameters (and hence lower contact resistance), with a wider range of dielectrics will be needed to fully understand their impact on the intrinsic transport properties of nanotube networks.

The studies discussed above do not explicitly consider defects in the employed nanotube networks that may be introduced during growth or processing. Recently however, the controlled low-level chemical functionalization of SWNTs has attracted considerable interest due to the possibility to tune and enhance their emission properties. By covalently grafting, *e.g.*, small aryl groups onto the nanotube sidewalls, so-called *sp*$^3$ defects are created, which act as efficient exciton traps and lead to red-shifted emission.[164,165] Measurements on single-nanotube devices indicated a decrease of the SWNT conductivity upon *sp*$^3$ functionalization,[166,167] however, the impact of *sp*$^3$ defects on charge transport in random nanotube networks was only recently investigated for pristine and functionalized (6,5) SWNT network FETs.[152] Both hole and electron mobilities decreased with increasing *sp*$^3$ defect density. Furthermore, a stronger temperature dependence of the carrier mobilities was found for functionalized nanotube networks compared to the unfunctionalized reference transistors (see **Fig. 5e**), with the differences being most pronounced in the low-temperature range. Due to the low number of *sp*$^3$ defects, it was assumed that they predominantly affected the intra-nanotube conductance within the networks rather than the inter-nanotube junctions, which may indicate that the variations of intrinsic nanotube mobility play a non-negligible role even in a network.[152]

**Thermoelectric measurements**

In the field of organic electronics, the study of thermoelectric properties and in particular the determination of the temperature- and carrier density-dependent Seebeck coefficients has led to remarkable new insights. The Seebeck coefficient can, for example, help to assess the relevance of electron–phonon coupling in molecular semiconductors. It is also a measure of the entropy transported by thermally excited charge carriers.[126,168,169] Doped semiconducting nanotube networks have been investigated for a while as possible thermoelectric materials (see above) and correlations between residual polymer, nanotube diameter and diameter distribution were made.[76,84,170] However, a deeper look may provide even more fundamental insights.

In a detailed study, Blackburn *et al.* investigated the impact of charge transfer doping on the temperature-dependent thermoelectric properties of semiconducting SWNT networks.[170] The electrical conductivity was found to be thermally activated, reflecting the inter-nanotube junctions as bottlenecks to charge transport especially in the low-doping regime. In contrast to that, these limitations to charge transport only had a minor impact on the Seebeck coefficient, which increased with temperature (100 K – 300 K) irrespective of the doping level of the SWNT network, exhibiting a behaviour similar to that of a highly conductive material. From the dependence of the Seebeck coefficient and consequently the power factor on the electrical conductivity, a transition between two different transport regimes was identified (see **Fig. 6a**). It was suggested that at low doping levels, charge transport was limited by the junctions between different SWNTs or bundles, whereas at high charge carrier densities, the intrinsic electronic properties (*i.e.*, acoustic phonon scattering) of the semiconducting SWNTs governed charge transport through the networks. The experimental data could indeed be fitted by the recently proposed semi-localized transport (SLoT) model that aims to combine hopping-like and metal-like transport as a function of temperature and carrier concentration in organic thermoelectric materials.[171]

Statz *et al.* performed temperature- and carrier density-dependent Seebeck measurements on monochiral (6,5) and large-diameter, multichiral semiconducting SWNT networks.[88] The obtained Seebeck coefficients were found to be electron-hole symmetric and decreased for increasing charge carrier densities. Indeed, the Seebeck coefficients of multichiral large-diameter and n-doped (6,5) nanotube networks followed the Heikes formula[172] remarkably well (see **Fig. 6b**), which would indicate a low-disorder system. Thermoelectric transport in the nanotube networks was further described according to the Boltzmann transport formalism, incorporating the framework of charge transport in heterogeneous media and the fluctuation-induced tunneling as corroborated by the extracted temperature-dependent carrier mobilities. Simulations according to this model suggested that one-dimensional scattering with acoustic and optical phonons is not sufficient to reproduce carrier scattering in the SWNT networks. Instead, scattering at inter-nanotube junctions needs to be included. Based on these results, trap-free SWNT networks with a narrow DOS distribution were proposed for optimized performance of electrical and thermoelectric devices.

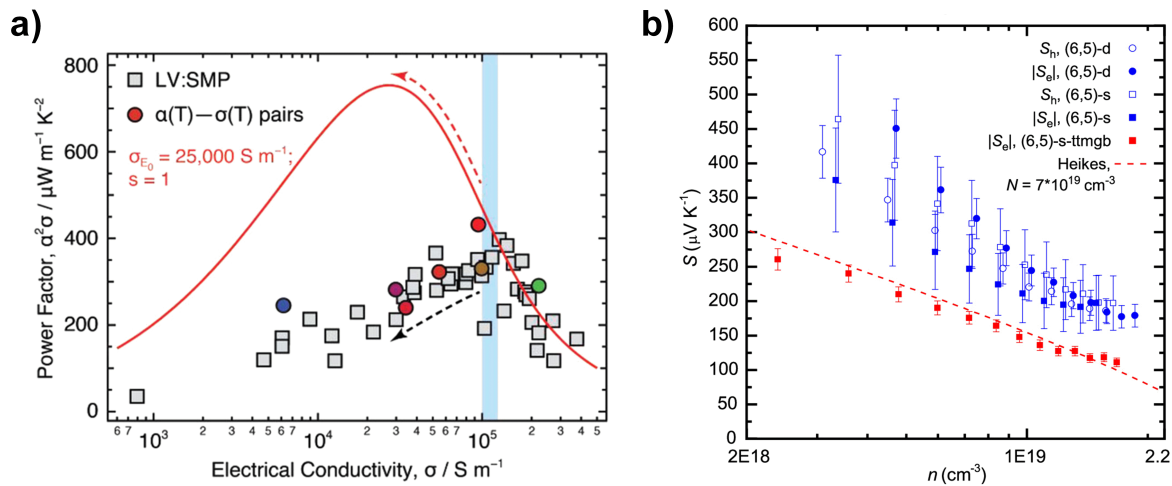

**FIG. 6.** Thermoelectric parameters of doped SWNT networks. **(a)** Power factor versus electrical conductivity for a semiconducting SWNT network (average diameter of ~1.35 nm, chemically doped). The red solid line corresponds to a simulation using a dispersive transport function. The vertical blue bar indicates the transition between two transport regimes. Reproduced with permission from Blackburn *et al.*, *Adv. Electron. Mater.* **2019**, *5*, 1800910.[170] Copyright 2019 WILEY-VCH. **(b)** Hole and absolute electron Seebeck coefficients ($S_h$ and $|S_e|$) of different (6,5) SWNT networks depending on the carrier density (electrostatically doped in a TG/BC FET). Blue circles are data points for a dense (6,5) nanotube network ((6,5)-d), whereas blue squares correspond to a sparse network ((6,5)-s). Red data points were measured on a sparse (6,5) SWNT network treated with the strong molecular n-type dopant ttmgb ((6,5)-s-ttmgb). The red dashed line is Heikes formula with $N = 7 \cdot 10^{19}$ cm$^{-3}$. Reproduced with permission from Statz *et al.*, *ACS Nano* **2020**, *14*, 15552-15565.[88] Copyright 2020 American Chemical Society.

Overall, temperature-dependent electrical and thermoelectric measurements have provided insights into the charge transport in semiconducting SWNT networks, although several aspects need to be considered in order to draw unambiguous conclusions. The progress in nanotube separation techniques finally allows for the preparation of purely semiconducting SWNT networks, however, the presence of excess wrapping polymer from the polymer-sorting process is expected to impede hopping of charge carriers between different nanotubes.[135] To probe the intrinsic network properties, the temperature dependence of the carrier injection has to be excluded, for example *via* four-point probe measurements. Complying with these demands, temperature-dependent measurements may enable conclusions on fundamental conduction mechanisms, such as the interplay between intra- and inter-nanotube transport in networks of different semiconducting SWNTs.[35,118,137] The nanotube junctions act as barriers to charge transport through the networks, and the activation energy required to promote a carrier across these junctions overall results in a thermally activated conduction mechanism in semiconducting SWNT networks as demonstrated in several studies. The temperature dependence of carrier mobilities in mixed semiconducting small-bandgap SWNT networks was lower than that in monochiral, large-bandgap nanotube networks.[137] Such insights are also important for the application of SWNT network transistors in electronic circuits with optimized performances. Indeed, for short channel transistors (channel length ~2 μm) with nanotube networks, the opposing trends of phonon scattering and thermally assisted hopping actually help to create ring-oscillators with a weak overall temperature dependence of the mobility and maximum oscillation frequency.[173]

Unfortunately, only the averaged overall network properties such as conductivities or mobilities can be extracted from temperature-dependent electrical measurements. Due to the lack of spatial resolution, the formation of certain transport pathways can only be assumed, but not visualized. This especially concerns networks consisting of multiple semiconducting SWNT species, in which preferential charge transport through large-diameter (*i.e.*, small-bandgap) nanotubes would be expected. To distinguish between different nanotube species, spectroscopic techniques can be utilized and will be discussed next.

**B. Spectroscopic techniques**

SWNTs exhibit strong and narrow absorption and emission features that are not only very specific to their structure but also highly sensitive to their environment and doping. The presence of charge carriers leads to substantial changes in the optical spectra of SWNTs. This property has enabled a wide range of methods utilizing the nanotubes' chirality-specific optical transitions as indicators or probes for charge carrier distribution and current flow in SWNT-based devices with spatial and energetic, *i.e.*, nanotube species, resolution.

The optical properties of SWNTs are governed by strongly bound electron-hole pairs, so-called excitons.[174,175] Due to the reduced dimensionality and thus lower dielectric screening, Coulomb interactions are drastically enhanced, which leads to exciton binding energies of several hundred meV.[176-179] Excitons can be either generated optically by photon absorption or by electrical excitation, usually electron-hole capture. The oscillator strength of nanotubes is quite large, giving rise to several narrow transitions ($E_{11}$, $E_{22}$ *etc.*) with high absorption coefficients[180,181] as shown in **Fig. 7a** for a dispersion of (6,5) SWNTs. Of the sixteen possible excitons only one is bright and can decay radiatively under photon emission in the near-IR (photoluminescence (PL) and electroluminescence (EL), respectively).[182-184] Emission occurs from the $E_{11}$ transition with minimal Stokes shift (few meV) and some red-shifted phonon sideband contribution[185] (PSB, see **Fig. 7a**). Due to their fast diffusion, excitons in SWNTs are able to sample large parts of the nanotube (several hundred nanometers) during their lifetime.[186,187] Hence, they are extremely sensitive to even small variations of the environment along the nanotube (*e.g.*, choice of solvent and surfactant coating, polymer-wrapping or adsorbed molecules)[188-190] and can be efficiently quenched through interaction with structural lattice defects including nanotube ends,[191] other excitons,[192] and charge carriers.[193] A detailed description and overview of the photophysics of SWNTs can be found in recent reviews on the topic.[183,194-196] Importantly, the spectral changes of absorption, photoluminescence and electroluminescence of SWNT networks can be employed to investigate the distribution of charge carriers and predominant transport paths in thin film devices even with some limited spatial resolution.

**Absorption spectroscopy**

Charge accumulation on SWNTs (*e.g.*, by electrostatic, electrochemical or chemical doping) is associated with the effective bleaching of absorption, which has found practical application in electrochromic devices with very thick SWNT layers.[59,61] While this effect has been the subject of various studies[107,197-199] and has for example enabled the quantitative analysis of doping levels,[200] charge-induced absorption bleaching has only recently been utilized to investigate charge transport in FETs with thin SWNT networks. Zorn *et al.* used charge modulation spectroscopy (CMS) as a method to study monochiral (6,5) and mixed semiconducting SWNT network transistors.[201] CMS has been applied widely to polymer[154,202-204] and small molecule organic semiconductor FETs[205,206] and helped to identify their charge carrier properties. It relies on the modulation of the optical transmission (*T*) of the semiconducting layer in an operating transistor upon application of a sinusoidal gate bias (on top of an offset voltage $V_{os}$) and hence modulation of the carrier density that follows a certain frequency (usually between 40 Hz and 10 kHz). The resulting differential change in transmission is recorded with a lock-in detection scheme (see **Fig. 7b**) and the detected signal (*ΔT/T*, positive for bleached absorption features and negative for new charge-induced transitions) is attributed to mobile charge carriers, because trapped charges cannot follow the voltage modulation. Spatial resolution of the CMS signal can be achieved when a confocal microscopy setup is used to perform charge modulation microscopy (CMM).[204,207]

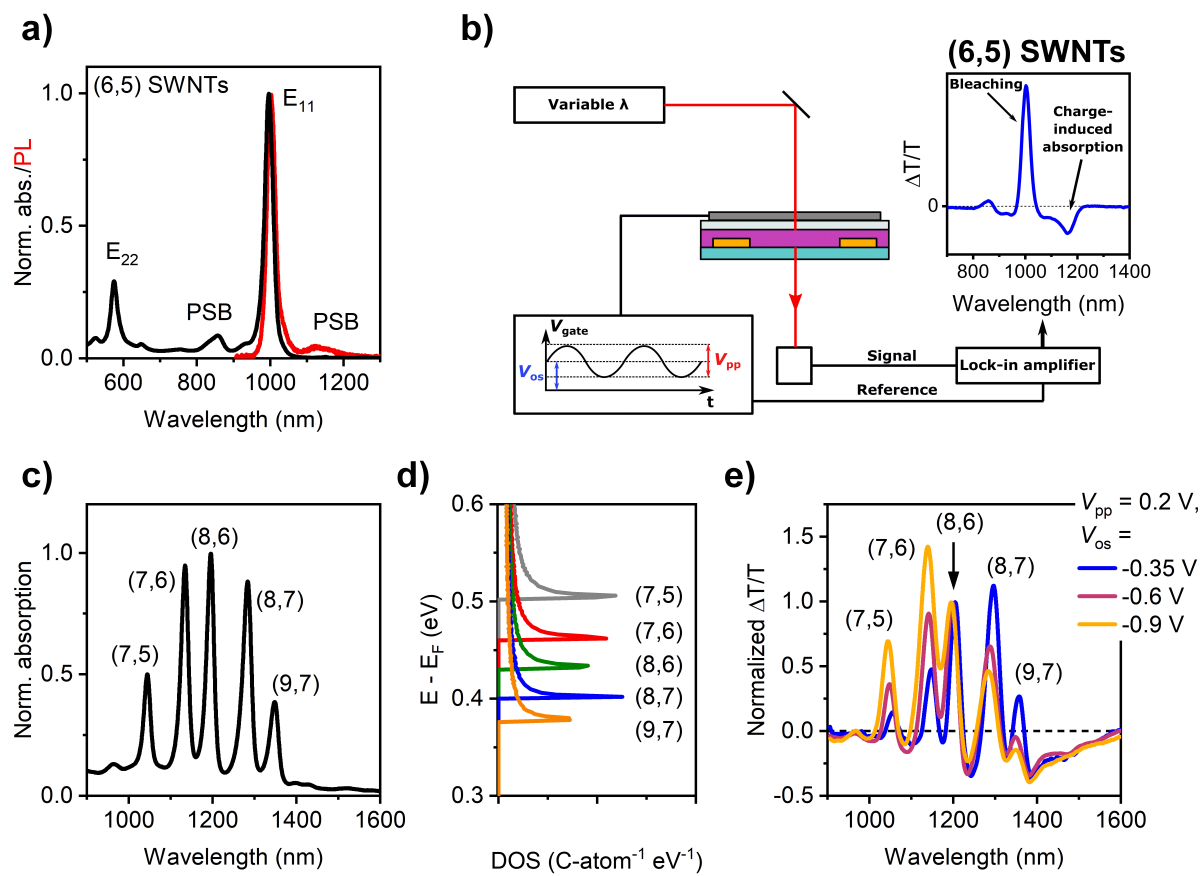

**FIG. 7.** Static and dynamic optical spectroscopy on SWNT networks. **(a)** Absorption spectrum (black) of a (6,5) SWNT dispersion in toluene and corresponding PL spectrum (red) after optical $E_{22}$ excitation at 575 nm. The main excitonic transitions ($E_{11}$, $E_{22}$) and the phonon sidebands (PSB) in absorption and emission are indicated. **(b)** Schematic illustration of the working principle of CMS. The charge density in the transistor channel is modulated with a sinusoidal gate bias and the optical response of the semiconducting layer is detected with a lock-in scheme. The corresponding spectrum was obtained on a monochiral (6,5) SWNT network FET. **(c)** Absorption spectrum of a multichiral SWNT dispersion containing five semiconducting species, and **(d)** first van Hove singularities of the conduction bands in the DOS of the different species. **(e)** Voltage-dependent CMS spectra (modulation frequency 363 Hz) of a multichiral SWNT network FET normalized to the signal of (8,6) SWNTs. Adapted under the terms of the Creative Commons BY-NC-ND 4.0 license from Zorn et al., *ACS Nano* **2020**, *14*, 2412-2423.[201] Copyright 2020 American Chemical Society.

The CMS spectrum of a monochiral (6,5) SWNT layer[201] showed gate voltage-dependent bleaching of the main $E_{11}$ absorption peak and red-shifted charge-induced absorption features (trions or polarons) as shown in **Fig. 7b**.[208,209] The correlation of the CMS signals of a multichiral network of five different semiconducting nanotubes (for absorption spectrum and DOS see **Fig. 7c,d**) with increasing offset voltage showed that at low carrier densities charges preferentially move through small-bandgap SWNTs (here (8,7) and (9,7)), irrespective of their comparatively low abundance in the network. With increasing total carrier concentration, the contribution of large-bandgap SWNTs (here (7,5)) to the overall charge transport increases (**Fig. 7e**).

**Photoluminescence spectroscopy**

Besides absorption bleaching, injected charges in SWNTs mediate non-radiative Auger recombination of excitons and hence PL quenching.[193,210] Owing to the high exciton and charge carrier mobilities in individual SWNTs but also in dense networks,[211,212] this process provides a very fast and efficient non-radiative decay channel even at low carrier concentrations and thus can be harnessed to study charge accumulation and transport in SWNT networks. Jakubka *et al.* used gate voltage-dependent static PL spectroscopy on multichiral semiconducting SWNT network transistors and found that charge carriers preferentially accumulated on small-bandgap SWNTs within the network, as their PL emission was the first to be quenched with increasing gate voltage (see **Fig. 8a,b**; similar network composition as in **Fig. 7c-e**).[42] They concluded that those SWNTs act as shallow traps for carriers of both polarities. Similar observations were made by Rother *et al.* who also calculated a gate voltage-dependent PL quenching factor for each chirality. These experimental results were nicely reproduced by an estimated Fermi-Dirac distribution of charge carriers among the different one-dimensional DOS (compare **Fig. 7d**).[117] However, as quenching of the PL intensity can arise from either trapped or mobile charge carriers, it cannot be correlated unambiguously with the actual charge transport properties of the network.

To address this issue, Zorn *et al.* combined voltage-dependent PL spectroscopy with modulation spectroscopy to perform charge-modulated PL measurements (CMPL, similar setup to CMS but with PL excitation and detection).[201] In these experiments, the differential PL quenching (*ΔPL*) induced by mobile charge carriers is recorded. Analogous to the CMS experiments in transmission, trapped charges do not contribute to the detected signal as they cannot follow the modulation frequency. From the fraction of quenched PL, which reflects the share of mobile charges on a given nanotube species, the chirality-dependent contributions within the SWNT network were assessed quantitatively (**Fig. 8c**). The obtained values were in good agreement with previously reported numerical simulations,[108] further corroborating the idea that at low carrier concentrations (*e.g.*, close to the subthreshold region) charge transport is dominated by the nanotubes with small bandgaps. Incidentally, these also exhibit the lowest injection barriers for both holes and electrons when using gold electrodes. As the carrier density increases, more of the larger-bandgap nanotubes start to contribute to charge transport until the entire network is involved at high gate voltages close to the maximum mobility. Importantly, these data suggest that the effective network density changes with applied gate voltage and that a monochiral network is preferable for more controlled and uniform device performance, *e.g.* with regard to a steeper and reproducible subthreshold slope.[213]

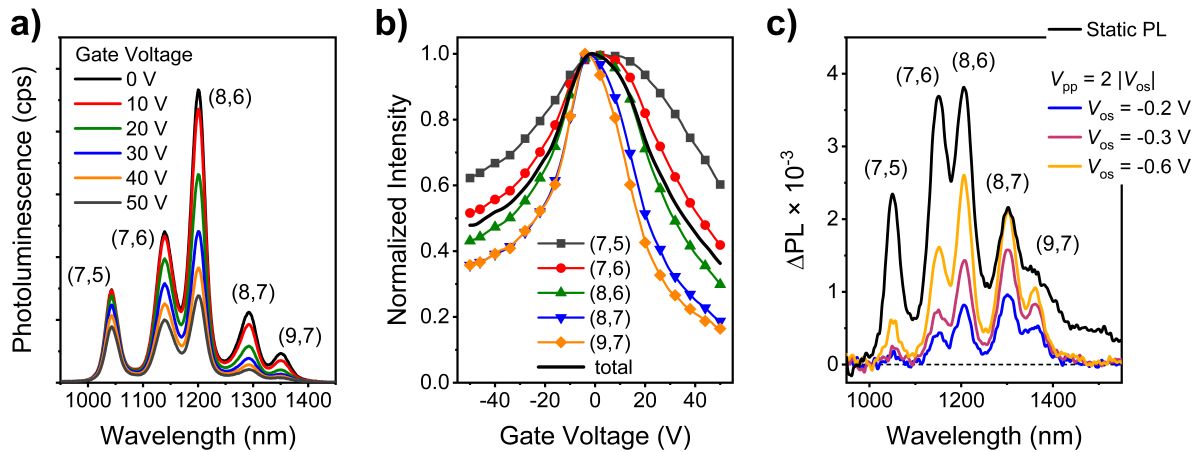

**FIG. 8.** Static and dynamic photoluminescence from gated SWNT networks. **(a)** Gate voltage-dependent PL spectra of a mixed semiconducting SWNT network FET. **(b)** Normalized intensities corresponding to the spectra in (a) show the chirality-dependent exciton quenching induced by static accumulation of charge carriers. **(a,b)** adapted with permission from Jakubka *et al.*, *ACS Nano* **2013**, *7*, 7428-7435.[42] Copyright 2013 American Chemical Society. **(c)** CMPL spectra of a multichiral semiconducting SWNT network FET at varying total charge densities (offset voltages $V_{os}$, modulation between off- and on-state of the device) show predominant quenching of emission from small-bandgap SWNTs through modulation of mobile carrier density. Adapted under the terms of the Creative Commons BY-NC-ND 4.0 license from Zorn *et al.*, *ACS Nano* **2020**, *14*, 2412-2423.[201] Copyright 2020 American Chemical Society.

**Electroluminescence imaging and spectroscopy**

Excitons can also be created electrically in SWNT FETs by impact excitation with strongly accelerated carriers (unipolar operation),[214,215] or by injection of carriers with opposite polarity and subsequent electron-hole capture (ambipolar operation, for a schematic illustration see **Fig. 9a**)[216,217] leading to EL in the near-IR. Here we will mainly consider EL caused by ambipolar carrier injection and capture as demonstrated for various aligned and random SWNT networks.[41,42,117,218,219] Early on, Adam *et al.* reported on mainly p-type SWNT network FETs fabricated from different nanotube raw materials that showed EL occurring close to the drain electrode.[220] They found that the emission spectra substantially differed from the corresponding absorption spectra of the nanotube networks, as the emission maxima appeared significantly red-shifted. This predominant EL contribution of large-diameter SWNTs was attributed to a higher current density in these nanotubes, which was corroborated by a simulation of the EL spectrum. At the same time Engel *et al.* demonstrated EL in more clearly ambipolar FETs based on aligned SWNTs with diameters in the range of 1.3 to 1.7 nm that were sorted by density gradient ultracentrifugation.[218] They also found a red-shift of the emission maximum (~1950 nm) of the EL and PL spectra as well as spectral narrowing compared to the expected emission distribution (see **Fig. 9b**), which they assigned to predominant current flow through large-diameter nanotubes and exciton transfer from large-bandgap to small-bandgap SWNTs. These initial studies suggested that EL spectra might be a good indicator for preferential charge transport in nanotube networks but that energy transfer between nanotubes may play an important role as well as seen for the PL spectra.[211,212] Networks with fewer nanotube species and with well-distinguishable emission features were clearly desirable to improve interpretation together with possible spatial resolution of emission.

Semi-aligned SWNT networks with only five distinct nanotube species selected *via* polymer-wrapping were later studied by Rother *et al*. to determine preferential EL compared to PL.[117] By recording EL spectra at different gate voltages and calculating the share of emission for each chirality, the chirality and carrier concentration dependence of charge transport in such multichiral semiconducting SWNT networks was analyzed quantitatively (**Fig. 9c,d**). In agreement with CMS and CMPL data (see above), the nanotubes with the smallest bandgaps were found to contribute predominantly to the EL and thus current, even though these species only accounted for a small portion of all nanotubes within the network. This was evident from their large EL shares, which were significantly higher than what would be expected from their nominal abundance and PL spectra recorded from the same channel region. Nanotubes with large bandgaps barely contributed to charge transport at low gate voltages and only started to show EL at higher carrier densities. This notion was further corroborated by numerical simulations of current and EL distribution depending on gate voltage in a numerically simulated network with the same composition.[108]

Similar to single nanotube devices,[175,217] EL of SWNT network FETs in the ambipolar regime occurs from a narrow recombination and emission zone (~1 μm observable width[117]) within the transistor channel (see **Fig. 9a**). As previously demonstrated for ambipolar polymer, small organic molecule and quantum dot transistors,[221-223] the emission zone can be moved across the entire channel depending on the applied voltages. Composite EL images can be created by combining all emission images of a voltage sweep (ideally for a constant current), which enables a genuine visualization of the total EL and - based on the required carrier continuity - also of current pathways.[224] Since all injected holes and electrons have to recombine, the EL in a certain area indirectly reflects the corresponding current density, assuming that the PL quantum yield is uniform. The spatial resolution, however, is limited by the optical setup and the emission wavelength.

Malhofer *et al.* combined such EL mapping with electrical transistor measurements to study the percolation behavior of nearly monochiral (6,5) SWNT in a polymer (PFO-BPy) matrix.[225] Using very low nanotube concentrations close to the percolation threshold, they were able to map only a few distinct percolation paths, whereas the majority of SWNTs did not participate in charge transport as shown by comparison with PL images. Remarkably, the visualization of single current pathways revealed that these conductive paths are far from uniform and can even fork into several trails. With increasing SWNT concentrations, the number of transport pathways also increased but remained clustered rather than being homogeneously distributed. This non-uniformity can be attributed to inhomogeneities arising from the network deposition, but also to the presence of some minority species with smaller bandgaps that should appear brighter in EL images.

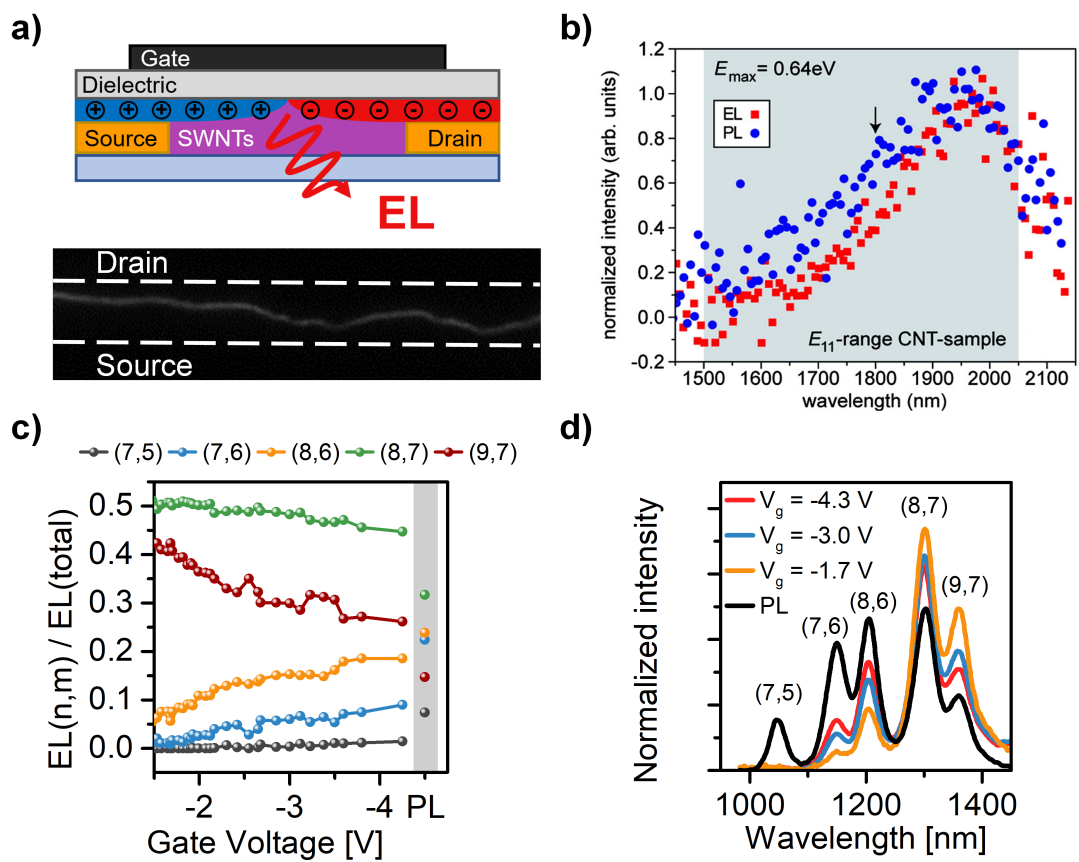

**FIG. 9.** Electroluminescence from SWNT networks. **(a)** Schematic illustration of electroluminescence from a FET biased in the ambipolar regime. A recombination and emission zone is formed within the channel where electron and hole accumulation layers meet (electrode distance 20 µm). **(b)** PL (excitation wavelength 514.5 nm) and EL spectra of SWNT FETs, showing red-shifted, narrower emission compared to the expected distribution (shaded in grey). Reproduced with permission from Engel et al., ACS Nano **2008**, 2, 2445-2452.[218] Copyright 2008 American Chemical Society. **(c)** EL share of individual nanotube species in a mixed semiconducting SWNT network depending on the total charge carrier density (gate voltage), and **(d)** corresponding EL spectra in comparison with the PL spectrum of the SWNT network. **(c,d)** reproduced with permission from Rother et al., ACS Appl. Mater. Interfaces **2016**, 8, 5571-5579.[117] Copyright 2016 American Chemical Society.

Spatially resolved EL was also used to investigate charge transport in multichiral SWNT networks. Jakubka et al. demonstrated electroluminescence from FETs with five different polymer-sorted semiconducting nanotube species (as in **Fig. 8a**).[42] For such a random, multichiral SWNT network, it is expected that conductive paths with lower and higher resistance are formed as a consequence of different nanotube species but also variations in nanotube density due to the deposition method (here drop-casting). This was reflected in a very non-uniform emission as shown in the composite EL images in **Fig. 10a**. Notably, the observable current pathways were determined by the assignment of the source and drain electrodes, which the authors attributed to variations in energy level alignment at the contacts due the different SWNT diameters, and thus preferential injection of either electrons or holes.

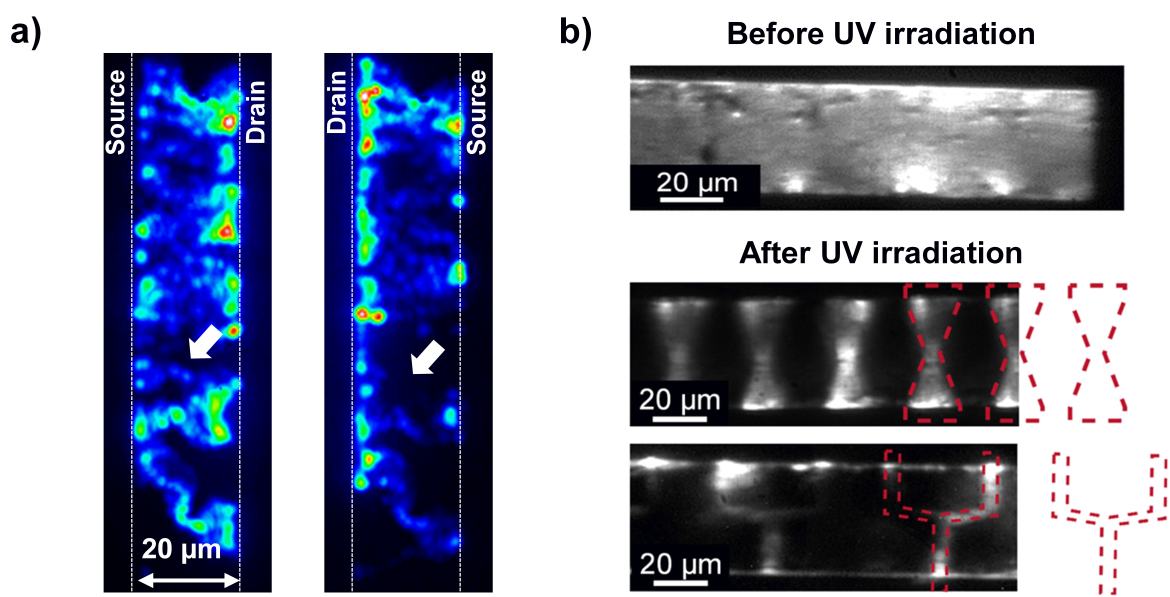

**FIG. 10.** Visualization of charge transport by electroluminescence imaging. **(a)** Composite EL images of a multichiral semiconducting SWNT network FET ($L = 20$ μm) with different hole and electron-injecting electrodes (exchanged source and drain). The white arrows mark emission spots that only appear when electrons are injected from the right electrode (see left image), but not from the left electrode (see right image). Reproduced with permission from Jakubka *et al.*, *ACS Nano* **2013**, *7*, 7428-7435.[42] Copyright 2013 American Chemical Society. **(b)** Composite EL images of (6,5) SWNT network FETs with photoswitchable spiropyran molecules embedded in the polymer dielectric. Upon UV irradiation through photomasks (red dashed outlines), the spiropyran locally switches to its merocyanine isomer, creating patterns with high (not exposed) and low (exposed) carrier mobilities. Reproduced with permission from Brohmann *et al.*, *ACS Appl. Mater. Interfaces* **2020**, *12*, 28392-28403.[226] Copyright 2020 American Chemical Society.

The notion of the composite EL image reflecting the predominant current pathways in a carbon nanotube network was further demonstrated by actually patterning such current pathways in an operating device as shown by Brohmann *et al.*[226] To achieve this, the authors incorporated photoswitchable spiropyran molecules into the polymeric gate dielectric layer of their (6,5) SWNT network FETs. Upon UV irradiation, the spiropyran undergoes a ring opening to form its merocyanine isomer, which was found to significantly reduce both hole and electron mobilities but did not affect the PL yield. Without any treatment the composite EL image of the channel was very homogeneous, representative of the uniform charge transport through a dense network of monochiral nanotubes. After UV light exposure through photolithography masks, regions of low (*i.e.*, exposed) and high (*i.e.*, not exposed) carrier mobilities were formed within the SWNT network. This patterning was confirmed by composite EL images that reproduced the mask structures (see **Fig. 10b**) while PL images remained uniform. The introduction of high and low mobilities areas even enabled a certain degree of deviation of the current pathways from the direction of the lateral electric field between source and drain electrodes.

The characterization of SWNT-based FETs using electro-optical techniques has substantially contributed to an improved understanding of the fundamental processes of charge accumulation and transport in SWNT networks. Especially for networks consisting of different SWNT species, the utilization of chirality-characteristic optical transitions and their spectral responses to the presence of charge carriers provided detailed insights. Charge carriers in multichiral semiconducting SWNT networks preferentially accumulate on and move through those nanotubes with the smallest bandgaps at low carrier concentrations (in the subthreshold region), whereas large-bandgap species only contribute at higher carrier concentration (in full accumulation). Hence, the performance parameters of multichiral semiconducting SWNT network FETs such as subthreshold slope, current on/off ratios, maximum transconductance and linear carrier mobilities strongly depend on the network composition.[117] These insights raise new questions about the validity of commonly used models and equations to describe the current-voltage characteristics of SWNT network FETs in the low-voltage regime[201] or charge transport mechanisms on the microscale.

The investigation of SWNT networks with techniques combining spectral and spatial resolution remains a challenging task, but is highly desirable to obtain further insights into the relationship of charge transport properties and microscopic network structure. Although EL mapping has enabled the visualization of preferential current pathways and individual percolation paths, the achievable spatial resolution on the order of ~1 µm prevents studies on a single nanotube level. Furthermore, the role of diameter-dependent junction resistance versus intra-nanotube transport cannot be resolved by these measurements. For understanding transport on a microscopic level, scanning probe techniques should be considered.

## C. Scanning probe techniques

Studying charge transport processes in SWNT networks on a microscopic level requires the use of methods that are capable of providing extremely high spatial resolution. In this context, scanning probe techniques such as conductive atomic force microscopy (c-AFM) or Kelvin probe force microscopy (KPFM) have proven to be very useful, as they are able to resolve current pathways on a nanometer to micrometer scale. Key parameters such as the conductance and resistance can be determined locally, *i.e.*, along individual nanotubes or at specific nanotube-nanotube junctions.

**Conductive atomic force microscopy (c-AFM)**

Conductive AFM is a frequently used technique to investigate conduction paths in organic semiconductors such as polymers,[227] small molecules,[228,229] or blends thereof.[230] Usually, the semiconductor is deposited on an insulating substrate but in electrical contact with an electrode (vertically or laterally). A conductive AFM tip (usually Pt- or Pt/Ir-coated) is used to raster scan over the sample in contact or intermittent contact mode while a fixed bias is applied between the sample and the tip. This enables the simultaneous measurement of the surface topography and current flow between electrode and tip at each position. The experimental setup for c-AFM measurements is schematically illustrated in **Fig. 11a** including a current map of a network of polymer-sorted (6,5) nanotubes in contact with a gold electrode measured in air. Note that reproducible c-AFM images of nanotubes and their networks are very challenging due to the need to be in contact with the sample while scanning without altering the surface (*e.g.*, moving the nanotubes) nor damaging the cantilever tip (abrasion of the metal coating). Scanning with an intermittent contact mode is more suitable for such samples.[231,232]

Early reports demonstrated the capabilities of c-AFM to locally probe the electrical properties of individual multi-walled and single-walled carbon nanotubes, albeit mostly in the form of bundles.[233-238] For example, Dai *et al.* measured a linear increase of the resistance along a multi-walled carbon nanotube with distance from the Au contact by c-AFM.[236] Through acquisition of current-voltage (*I-V*) curves at different points along SWNTs, de Pablo *et al.* found a non-linear dependence of the resistance on the nanotube length, which they attributed to elastic scattering of electrons.[237] Stadermann *et al.* reported an exponential decay of the local conductance of SWNTs in close proximity to nanotube junctions.[238] Note that at the time of these early measurements, neither the type (metallic or semiconducting) nor the precise diameter of the investigated nanotubes could be determined.

Conductive AFM was also applied to measure the resistance along nanotube segments and across nanotube-nanotube junctions in SWNT networks. However, most of these studies used random networks consisting of both semiconducting and metallic nanotubes. Stadermann *et al.* acquired conductance images on networks of as-grown nanotubes on Si/SiO$_2$ substrates with different degrees of interconnectivity.[239] The conductance of specific regions within networks of high interconnectivity increased with the number of connecting pathways to the electrode. This lower resistance was the result of the formation of parallel circuits and thus a higher probability of direct paths with only metallic, low-resistance SWNTs. For nanotube networks with few interconnections, the conductance drastically decreased at specific points that coincided with SWNT crossings, which were attributed to junctions between semiconducting and metallic nanotubes. By back-gating the sample with a positive bias, the current flow through semiconducting nanotubes was suppressed and only the metallic paths were visualized. Nanotube junctions had been previously investigated by Fuhrer *et al.* who demonstrated the formation of Schottky barriers between semiconducting and metallic SWNTs.[111] Clearly the precise nature of the junctions in a network plays an important role.

A detailed study on charge transport in SWNT networks using c-AFM was provided by Nirmalraj *et al.*[110] Their networks consisted of arc-discharge nanotubes (mixed metallic and semiconducting) with diameters in the range of 1.2 – 1.8 nm and bundles thereof. Individual junctions were identified from current maps and the resistance was measured along the possible current pathways, revealing steep jumps in resistance at the nanotube junctions on the order of 100 kΩ as shown in **Fig. 11b**. Furthermore, the conductivity along a single nanotube or bundle was calculated from the slope of the resistance versus the position. A detailed analysis based on the diameters of the involved nanotubes revealed that the measured junction resistance depended on the diameter of the crossing nanotubes and bundles. A mean resistance of 98 kΩ for junctions between individual SWNTs was calculated, whereas the junction resistances between SWNTs and bundles as well as between bundles of different sizes were significantly higher. These findings emphasized the negative effect of bundles on the conductivity in nanotube networks.

Furthermore, the dependence of junction resistance on doping was analysed. Treatment with nitric acid, which leads to p-doping,[240] generally reduced the junction resistance, with a threefold drop for junctions of individual SWNTs. Annealing (*i.e.*, de-doping) of the network increased the junction resistance again. Both treatments were shown to predominantly affect the junction resistance and not the nanotube conductivity, corroborating the notion of the overall SWNT network performance being dominated by the junctions.

The effect of doping on the resistance of SWNTs and their junctions was further investigated by Znidarsic *et al.*[109] From local *IV*-curves on chemical vapor deposition (CVD)-grown SWNTs (metallic and semiconducting) and bundles with different diameters, the nanotube resistances were calculated to be 3 – 16 k$\Omega$/µm, and the junction resistances to range between 29 and 532 k$\Omega$. They demonstrated that the junction resistances decreased with increasing diameter of the crossing SWNTs as expected from theory, which was however in contrast to the results of Nirmalraj *et al.* (see above).[110] These data sets should be compared with caution due to the differences between the studied samples (*e.g.*, different nanotube growth and deposition processes, presence or absence of surfactants). Doping of the network *via* treatment with nitric acid, however, again reduced the junction resistances by approximately a factor of 3, whereas the nanotube resistances did not change significantly.

The previous two studies focussed on the mean junction resistance in the network. Recently, Stern *et al.* studied the change in resistance of individual junctions upon doping with nitric acid.[241] They found that the exposure to nitric acid improved the conductance of each investigated junction and suggested that the underlying causes were charge transfer doping and degradation of the surfactant. Note also, that in all of the above cases the networks contained both metallic and semiconducting nanotubes. While the conductance of metallic nanotubes should not change with doping, that of semiconducting nanotubes should increase by several orders of magnitude.

So far only one report in the literature has focussed on c-AFM of purely semiconducting nanotube networks with known composition. Bottacchi *et al.* investigated the conduction paths in polymer-sorted networks of (7,5) SWNTs, which were integrated in bottom-gate/bottom-contact FETs and characterized under inert conditions to avoid doping of the nanotubes in air.[242] Current maps of the network were obtained, which enabled the calculation of in-plane resistivities perpendicular and parallel to the electrodes. Both values were very similar, indicating an isotropic distribution of nanotubes as expected for a random, spin-coated network. Although the networks under investigation were already integrated in a transistor structure, the authors did not show further data, *e.g.*, dependence of the network conductance on the electrostatic doping level induced by a gate voltage or individual junction resistances.

**KPFM and related techniques**

Another prominent method among the scanning probe techniques is KPFM, which allows the surface potential or work function of a surface to be mapped with high resolution.[243] In KPFM, a metallic cantilever is scanned over the (semi-)conducting sample under investigation and the potential difference is recorded. The method was successfully employed to elucidate charge injection and charge transport in organic thin film transistors.[244-247] In studies on individual SWNTs, this method has provided detailed insights into local electrostatic properties and conduction mechanisms.[248,249] For example, Fuller *et al.* used modified KPFM to obtain the potential profiles along individual nanotubes in transistor structures for different source-drain voltages.[250] The same technique was used in a subsequent study to determine the mean free path of carriers in SWNTs.[251]

Using KPFM on SWNT networks enables spatial mapping of the potential. Okigawa *et al.* recorded the potential across the channel in FETs based on dense, CVD-grown nanotube networks.[252] In the subthreshold regime, they found a constant potential at zero source-drain bias, whereas the potential profile for a small source-drain voltage exhibited step-like drops. A non-uniform distribution of regions with higher and lower resistance appeared in the corresponding resistance distribution maps acquired by c-AFM. The data indicated that the current flow was restricted to only a few conductive paths connecting areas of low resistance, meaning only few nanotubes contributed to the charge transport in the subthreshold regime. In the on-state of the device, however, KPFM images revealed a smooth potential gradient along the direction of the channel, suggesting that the current is more evenly distributed over the nanotube network at higher carrier densities. The authors attributed these observations to the gate voltage-dependent resistance of the semiconducting SWNTs.

Recently, Hao *et al.* used KPFM to spatially map the surface potential in an electronic ratchet device based on a semiconducting SWNT network FET where the nanotubes were chemically p-doped.[63] Comparing the potential profiles before and after application of a voltage stress (-15 V) to the device indicated a decrease in the potential drop for hole injection from the source electrode, whereas the potential drop increased at the drain electrode. This observation of different injection barriers was ascribed to the redistribution of dopant counterions within the transistor channel during the stressing process, inducing a spatial asymmetry in the device and thus explaining the observed rectifying behaviour.

Closely related to KPFM is electrostatic force microscopy (EFM), which directly records the electrostatic force acting on the cantilever. Topinka *et al.* used EFM to study SWNT transistors with mixed metallic/semiconducting CVD-grown nanotube networks.[253] They observed pronounced differences between individual devices. In some samples the network resistance was relatively low and the voltage dropped smoothly between the source and drain electrodes, for others the network resistance was high with abrupt voltage drops at specific points within the channel (see **Fig. 11c**). Simulations of current flow in random nanotube networks indicated that the presence or absence of steep voltage drops could be the result of single percolation paths at certain points in the film. Substantial variations in device performances probably resulted from differences in the ratio of semiconducting to metallic nanotubes. Consequently, the authors noted that enrichment of semiconducting SWNTs was a key step towards reliable device measurements.

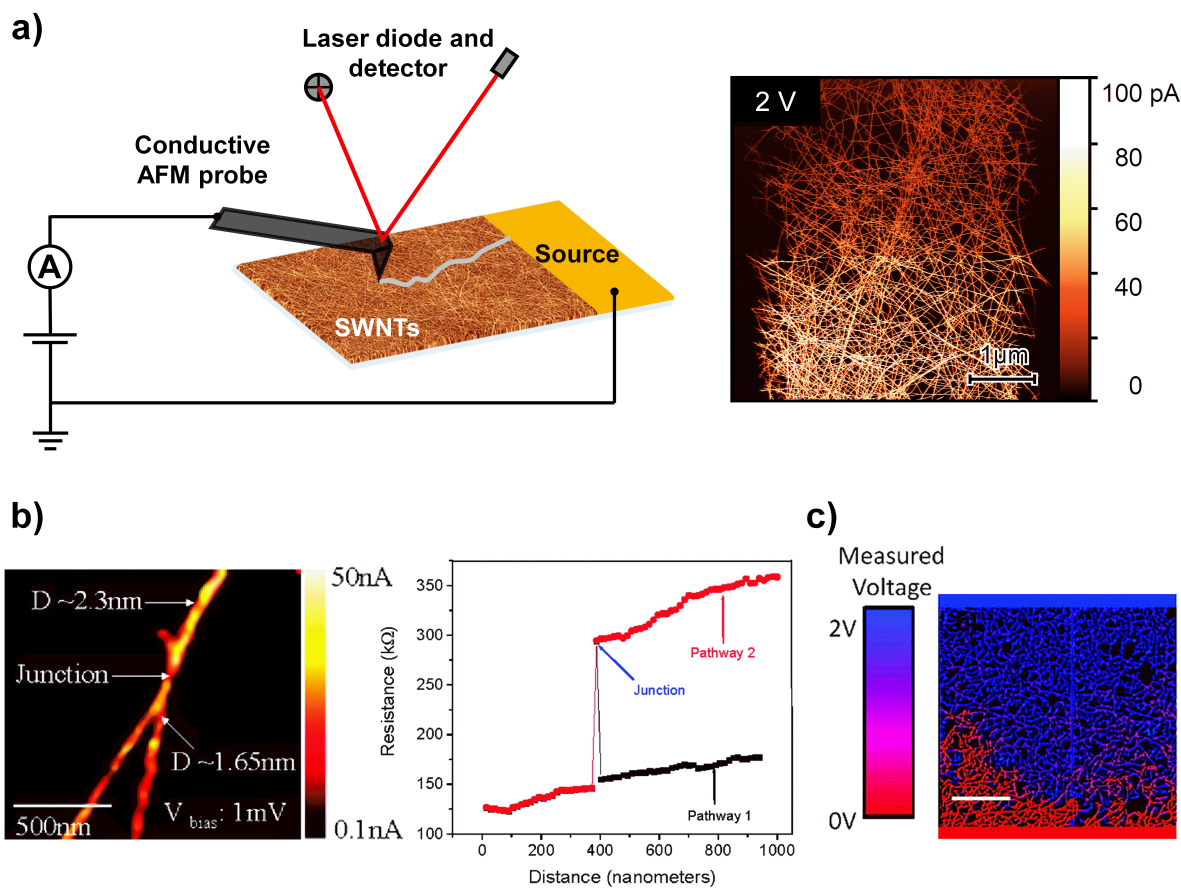

**FIG. 11.** Scanning probe techniques. **(a)** Schematic illustration of the working principle of conductive AFM (c-AFM) and c-AFM image of a (6,5) SWNT network in contact with a gold electrode located below the image (bias voltage, 2V; scale bar, 1μm; image courtesy of Constantin Tormann, Heidelberg University). **(b)** Current map showing a nanotube junction (left; electrode position is above the image) and corresponding resistance analysis (right) measured with c-AFM. Pathway 1 corresponds to the current trace along the ~2.3 nm SWNT bundle, whereas pathway 2 follows the ~1.65 nm SWNT after the junction. Reproduced with permission from Nirmalraj *et al., Nano Lett.* **2009**, *9*, 3890-3895.[110] Copyright 2009 American Chemical Society. **(c)** Electrostatic force microscopy (EFM) image (40 × 40 μm, scale bar 10 μm, source and drain electrodes located at the bottom and top of the image) of a SWNT network showing a steep voltage drop near the bottom electrode. Reproduced with permission from Topinka *et al., Nano Lett.* **2009**, *9*, 1866-1871.[253] Copyright 2009 American Chemical Society.

Microwave impedance microscopy (MIM) is yet another scanning probe technique that was recently applied to SWNT networks by Seabron *et al*.[254,255] In MIM, the sample is scanned with a cantilever and the spatially resolved microwave reflectivity is detected through the mismatch of sample and tip impedance. Combined with AFM, this technique is capable of directly probing the electronic properties of SWNTs in aligned arrays and random networks, *i.e.*, distinguish between metallic and semiconducting nanotubes. Incorporating a voltage modulation (microwave impedance modulation microscopy) even enables tracing the nanotubes' DOS similar to scanning tunnelling spectroscopy.[254]

In summary, scanning probe techniques have been successfully applied to investigate charge transport in different SWNT networks. Although in general very challenging, they have enabled the visualization of conduction paths and maps of current distribution. The key role of nanotube junctions for the overall network performance was confirmed. Junction resistances were on the order of 100 k$\Omega$ and intra-tube resistances around 10 k$\Omega$/µm.[109,110] Doping of SWNT networks, *i.e.* increasing the number of mobile charge carriers, generally caused a decrease in junction resistance and hence an increase in nanotube network conductivity.[109,110,241] In networks consisting of metallic and semiconducting nanotubes, the ratio between these different types led to the formation of very different conducting paths as predicted by percolation theory.[253] Overall, these reports suggest that charge transport in SWNT networks is strongly dominated by the junctions and any residual metallic nanotubes. Yet, the discussed techniques require specific knowledge about the system as they can neither distinguish between different nanotube species nor clearly between metallic and semiconducting SWNTs (except MIM). Most studies used networks of as-grown nanotubes, which are very clean (no surfactants) and contain few bundles but also include nanotubes with a range of diameters and about 30% metallic nanotubes. Hence, the results are only of limited applicability to state-of-the-art SWNT networks with only semiconducting or even monochiral nanotubes.[34] Unfortunately, many questions regarding microscopic charge transport in purely semiconducting nanotube networks such as their junction versus intra-nanotube resistance depending on doping remain unanswered.

## V. CHALLENGES AND CONCLUSIONS

### A. Experimental and theory challenges

While there has been a tremendous amount of progress in the fabrication and application of purified semiconducting nanotube networks to the point where they can be used to produce fully integrated microprocessors,[12,13] there are still a lot of open questions regarding the details of charge transport within such networks. As shown above some aspects can be investigated in more detail now due to better control of the network density and composition. Electrical and thermoelectric measurements, dynamic absorption, photoluminescence and electroluminescence spectroscopies as well as scanning probe microscopies have been successfully applied to gain insight into the charge transport in carbon nanotube networks. However, a complete picture that enables targeted optimization will require further effort on the experimental and the theory side. Some of these challenges are summarized here.

Describing and understanding charge transport in SWNT networks requires theory and models on different levels. Firstly, the charge transfer between two overlapping semiconducting nanotubes of the same (large or small) or different diameters and with various orientations to each other must be modelled at a higher level of theory than what has been done so far.[112] Given the progress in computing power, even chiral nanotubes with relatively large unit cells should be accessible for modelling direct charge transfer similar to studies carried out for charge transfer between π-conjugated organic molecules.[256,257] While this may still not be representative of an actual contact it should provide basic information on diameter, carrier density, contact angle dependence *etc.* of the charge transfer probability.

Next, combining these basic parameters with the existing knowledge of transport within individual nanotubes should enable numerical modelling of entire networks with different densities and different compositions. Ideally these networks should not just be simplified to stick model in two dimensions but actually represent the three-dimensional structure of a network and include the corresponding electrostatics.[103] This task might be rather large and complex. Hence as a first step, information about the general current paths and temperature dependence of charge transport should already become accessible from 2D networks. The role of wrapping polymer or other surfactants in the form of additional tunneling barriers with certain energy levels might even be explored in a controlled manner and thus enable more direct comparison to experimental results.

Such numerical modelling of different types of networks and thus a deeper understanding of the impact of microscopic transport on macroscopic parameters such as apparent carrier mobilities of a thin film should enable the development of an analytical - albeit empirical - equation. Such an equation should be able to describe the temperature, carrier concentration and composition (diameter) dependence of a given semiconducting network and most importantly allow for physical insights to be drawn from the fitting parameters.

On the experimental side the challenges might be divided between experiments that would be useful to gain more insight into the transport physics and experiments for further optimization of device performance for applications. The former should of course inform the latter as well.

The question of residual wrapping polymer or surfactant of solution-processed networks and its potential impact on charge transport still remains open and requires comprehensive and careful experiments to be resolved. It is commonly assumed that excess polymer can act as a barrier to charge transport through SWNT networks, however, the problem lies within the many parameters that must be considered and cannot be easily controlled or even determined directly (*e.g.*, precise coverage and wrapping geometry of polymers on nanotubes after deposition). Furthermore, one has to distinguish between the polymer that is actually wrapped around (and likely strongly bound to) the nanotubes, and free, unbound polymer in solution, which is typically removed after SWNT network deposition simply by rinsing with a suitable solvent. Recent studies have suggested that the residual unbound polymer of spincoated semiconducting nanotube networks does not reduce the effective carrier mobility,[143] others have shown that the temperature dependence of transport can be affected.[142] With regard to the influence of residual polymer wrapped around the SWNTs, different approaches have been reported to remove it. One prominent example is the use of supramolecular wrapping polymers that can easily be cleaved and washed off after network deposition. Several studies reported a significant improvement in the performance of SWNT network FETs and thermoelectric devices upon removal of the polymer.[76,258,259] Another approach involved the removal of the wrapping polymer PFO-BPy through metal complexation with a rhenium salt, however, without significant impact on the transfer characteristics of aligned nanotube array transistors.[260] In contrast to that, the rhenium salt treatment of random (6,5) SWNT/PFO-BPy networks led to a higher degree of nanotube bundling and in fact a decrease in carrier mobilities for the resulting FETs.[35] These examples highlight the problem of trying to investigate one parameter affecting transport in a nanotube network without also changing another.

A perfectly clean network of well-separated and purely semiconducting nanotubes with known composition and deposited from the gas phase should be a good reference for solution-processed networks. Unfortunately, such samples are not yet available. Furthermore, while the role of the surrounding dielectric (metal oxides or polymers) was studied (see above) and showed a direct effect on the charge transport,[136,155,156] it is still unclear how to incorporate these effects into a universal model.

The macroscopic device parameters extracted from field-effect transistors and other electrical measurements give only limited insights into the microscopic charge transfer between nanotubes. More detailed scanning probe microscopy studies especially on networks of purely semiconducting and monochiral carbon nanotubes would be helpful. In particular, a quantitative analysis of the junction resistance between semiconducting nanotubes of a certain chirality, the intra-nanotube resistance and their dependence on doping (chemical or electrostatic) is necessary to be able to determine the main factors limiting transport in a given network. Finally, the impact of localized structural defects (by intentional covalent functionalization or unintentional intrinsic defects) needs to be understood on a microscopic level. Techniques that allow for the determination of microscopic and local carrier mobilities without the need of electrodes such as time-resolved THz spectroscopy[261,262] and time-resolved microwave conductivity measurements[263,264] should be very useful for that purpose.

For the commercial application of semiconducting nanotubes in circuits, a number of issues need to be resolved. The shelf-life of a nanotube ink is limited by the aggregation of the nanotubes in dispersion over time, leading to non-uniform networks with many bundles, aggregates and reduced device performance.[265] Achieving highly uniform nanotube networks processed from solution without extensive post-processing will be necessary for large-area nanotube circuits to become reality. The composition of a network must be precisely controllable and reproducible from batch to batch, not just in terms of absence of metallic nanotubes but also in terms of the length and diameter distribution of the semiconducting nanotubes. As demonstrated above even small amounts of small-bandgap nanotubes can affect the overall transport especially at lower carrier concentrations (*i.e.*, in the subthreshold region). Reducing this variability and energetic disorder of the network with a single nanotube chirality would also increase the effective carrier mobility. At the moment the only monochiral nanotube dispersions that are available on a larger scale and have been studied in field-effect transistors are (6,5) SWNTs with a relatively small diameter and hence larger injection barriers and lower intrinsic mobilities than large-diameter nanotubes.[35,151] Networks of large-diameter but still mixed semiconducting nanotubes result in the highest field-effect mobilities to date and are predominantly employed for circuit applications.[4,12,173] Hence, methods to produce large amounts of monochiral and large-diameter nanotubes should be pursued further. Recent progress in aqueous two-phase extraction has already enabled the purification and enrichment of several monochiral nanotube dispersions with diameters of around 1.41 nm.[266] Similar dispersions of large-diameter nanotubes might also be possible by polymer-wrapping.[267] It will be interesting to see where the true limits of carrier mobility and on/off current ratios for large-diameter/small-bandgap nanotube networks lie and whether they justify the effort of purification compared to multichiral semiconducting networks.

Small-bandgap nanotubes are more likely to show ambipolar transport even under ambient conditions. For ohmic contacts the maximum on/off current ratio for an ambipolar semiconductor is given by its bandgap.[268] The narrow bandgaps of large-diameter SWNTs would start to limit this ratio irrespective of the absence of metallic nanotubes. In any case, for low-power complementary circuits, controlled and selective hole or electron injection and transport are required. The predominant p-type behavior of many SWNT transistors is based on high work function contacts and electron trapping by oxygen and water.[27,28] For n-type transistors low work function electron-injecting contacts are often used,[8,29] while the controlled application of chemical dopants[30] or specific dielectrics[31] are also viable options. The use of alkaline and strongly reducing dopants may also help to remove water and electron traps that lead to hysteresis and poor subthreshold slopes.[32,151] Importantly, these approaches for creating purely p-type and n-type transistors with high on/off current ratios must be reproducible and stable under operating and ambient conditions.

As shown in the preceding sections, the precise composition of a nanotube network and possible residual minority species can have a large impact on the device characteristics. Reliable and simple methods and metrics to quantify the nanotube species in and the purity of dispersions and networks are required for characterization and quality control. While simple absorption spectroscopy can provide general information about the distribution of majority species even for samples with many chiralities through spectral fitting,[269] it is not sensitive enough at the very low percentages of minority species. Determining and controlling the composition of a nanotube network below chirality concentrations of 0.1% requires highly sensitive techniques. Photothermal deflection spectroscopy (PDS) on thin films can provide reliable measurements of absorbances several orders of magnitude below those of a good absorption spectrometer.[270] For example, minority nanotube species in thin films of (6,5) nanotubes as well as their Urbach tail could be identified with PDS.[88] However, this technique usually requires a custom setup and sufficiently thick films immersed in a high refractive index inert liquid.[126]

Photoluminescence excitation-emission maps can be highly sensitive and give quantitative distributions of the detected nanotube species[271,272] but PL spectroscopy is blind toward residual metallic nanotubes and those with very small bandgaps (< 0.8 eV) and thus beyond the typical detection limits (~1600 nm) for InGaAs detectors. The popular and easily accessible technique of Raman spectroscopy cannot provide quantitative information about the concentration of different SWNT species at all, as it is always limited by the employed laser wavelengths and the chirality-dependent variations of the scattering cross sections.[273-275] It can however be used to show the absence of certain species if suitable excitation lasers are used. More reliable identification techniques such as TEM[276] or electron diffraction[277] cannot be applied to large sample volumes. Hence, most studies will continue to rely on a combination of the above-mentioned techniques, keeping in mind their specific limitations.

To be able to compare different nanotube network devices and hence analyze differences and improvements, common standards for extracting and reporting device parameters especially carrier mobilities must be established. For examples, the effective gate capacitance should be measured directly on the device to ensure correct mobility calculations due to its dependence on the network density especially for very thin dielectrics.[23,26] The carrier mobilities in FETs with carbon nanotube networks strongly depend on the carrier concentration and hence gate voltage, which may lead to different numbers depending on the method of mobility extraction. The maximum field-effect mobility might be a comparable standard if a clear maximum is reached. However, one must ensure that this is indeed a mobility maximum and not an artefact of gate-modulated contact resistance, which has plagued mobility measurements of organic high-mobility semiconductors.[150,278] Ideally, contact resistance-corrected mobilities as extracted from gated four-point probe[149] or van der Pauw measurements[147,279] should be used. Given the high carrier mobilities in nanotube networks, even relatively low contact resistances can skew the extracted values and more importantly limit the maximum switching frequency of a transistor and the frequency of a ring-oscillator.[280] Moreover, the uniformity and reproducibility of device parameters is not only crucial for fundamental studies, but it directly determines the applicability in actual circuits. Thus, average mobility values with standard deviations for a statistically relevant number of devices instead of values from "hero devices" should always be reported.

The large surface area of carbon nanotubes and the assumed direct impact of their environment on charge transport provides ample opportunities for chemo- and bio-sensing[64-66,69,167,281] as well as (optical) memory applications.[57,282-284] Functionalization (covalent and non-covalent) of nanotubes has already been applied for a range of devices, although the underlying mechanisms are not always clear. A more fundamental understanding of charge transport in networks and the parameters that determine high or low mobility, threshold shifts *etc*. would be very helpful to further improve the performance of such sensors. Finally, the peculiarities of charge transport in nanotube networks (especially controlled hysteresis) may also enable their application in future neuromorphic devices.[285-288]

## B. Conclusions

Understanding and controlling charge transport in networks of single-walled carbon nanotubes has vastly improved over the last two decades. At first, researchers had to deal with and were frustrated by poorly controlled networks with many metallic SWNTs or semiconducting nanotubes of unknown diameter. These samples showed poor current modulation, large hysteresis and insufficient reproducibility with hardly any perspective for application in high-performance electronics. Today the synthesis and purification of semiconducting and even monochiral nanotubes enables their use as a solution-processable electronic material with specific and tunable properties that is competitive with or even better than many other semiconductors for thin film transistors and other (opto-)electronic devices. Not only is it now possible to investigate the intrinsic charge transport properties of networks with specific and well-controlled compositions, they can be applied in integrated circuits on an industrial scale. While there are still many open questions, the groundwork for answering them has been laid by the new SWNT purification and sorting techniques as well as the wide range of available characterization and simulation methods.

## ACKNOWLEDGMENTS

The authors acknowledge funding from the European Research Council (ERC) under the European Union's Horizon 2020 research and innovation programme (Grant agreement No. 817494 "TRIFECTs").

**DATA AVAILABLITY:** As this is a review, the data supporting the various figures can be found in the corresponding referenced publications. All other data is available on request from the authors.